\documentclass[reprint,amsmath,amssymb,aps,prd,onecolumn,nofootinbib,superscriptaddress]{revtex4-1}
\usepackage{graphicx}
\graphicspath{{fig/}} 
\usepackage[letterpaper,margin=1in]{geometry}
\usepackage{booktabs}
\usepackage{multirow}
\usepackage{pifont}
\usepackage{todonotes}
\usepackage{amssymb}
\usepackage{amsmath}
\usepackage{xcolor, colortbl}
\usepackage{array}
\usepackage[colorlinks = true, bookmarksopenlevel=4,
            linkcolor = blue,
            urlcolor  = blue,
            citecolor = blue,
            anchorcolor = blue]{hyperref}

\newcommand\pubnumber{arXiv: }
\newcommand\pubdate{\today}

\newcommand\pubblock{\rightline{\begin{tabular}{l} \pubnumber\\
         \pubdate \end{tabular}}}
\newenvironment{Abstract}{\begin{quotation} \begin{center}
                       ABSTRACT
     \end{center}\bigskip  }{\end{quotation}}

\newcommand\snowmass{\begin{center}\rule[-0.2in]{\hsize}{0.01in}\\\rule{\hsize}{0.01in}\\
\vskip 0.1in Submitted to the Proceedings of the US Community Study\\
on the Future of Particle Physics (Snowmass 2021)\\
\rule{\hsize}{0.01in}\\\rule[+0.2in]{\hsize}{0.01in} \end{center}}

\begin{document}

\pubblock

\title{Physics Opportunities in the ORNL Spallation Neutron Source Second Target Station Era}

\bigskip

\newcommand{\UTAdesc}{\affiliation{Department of Physics, University of Texas at Arlington, Arlington, TX, 76019, USA}}
\newcommand{\Dukedesc}{\affiliation{Department of Physics, Duke University, Durham, NC, 27708, USA}}
\newcommand{\TUNLdesc}{\affiliation{Triangle Universities Nuclear Laboratory, Durham, NC, 27708, USA}}
\newcommand{\FNALdesc}{\affiliation{Fermi National Accelerator Laboratory, Batavia, IL, 60510, USA}}
\newcommand{\UTKdesc}{\affiliation{Department of Physics and Astronomy, University of Tennessee, Knoxville, TN, 37996, USA}}
\newcommand{\ORNLdesc}{\affiliation{Oak Ridge National Laboratory, Oak Ridge, TN, 37831, USA}}
\newcommand{\IGFAEdesc}{\affiliation{Instituto Galego de Fisica de Altas Enerxias, A Coru\~{n}a, Spain}}
\newcommand{\NCSUdesc}{\affiliation{Department of Physics, North Carolina State University, Raleigh, NC, 27695, USA}}
\newcommand{\USDdesc}{\affiliation{Physics Department, University of South Dakota, Vermillion, SD, 57069, USA}}
\newcommand{\NCCUdesc}{\affiliation{Department of Mathematics and Physics, North Carolina Central University, Durham, NC, 27707, USA}}
\newcommand{\CMUdesc}{\affiliation{Department of Physics, Carnegie Mellon University, Pittsburgh, PA, 15213, USA}}
\newcommand{\IUdesc}{\affiliation{Department of Physics, Indiana University, Bloomington, IN, 47405, USA}}
\newcommand{\TAMUdesc}{\affiliation{Department of Physics and Astronomy:Mitchell Institute for Fundamental Physics and Astronomy, Texas A\&M University, College Station, TX, 77840, USA}}
\newcommand{\SLACdesc}{\affiliation{SLAC National Accelerator Laboratory, Menlo Park, CA, 94025, USA}}
\newcommand{\Hawaiidesc}{\affiliation{Department of Physics and Astronomy, University of Hawaii, Honolulu, HI, 96822, USA}}
\newcommand{\Tuftsdesc}{\affiliation{Department of Physics and Astronomy, Tufts University, Medford, MA, 02155, USA}}
\author{J.~Asaadi}\UTAdesc
\author{P.S.~Barbeau}\Dukedesc\TUNLdesc
\author{B.~Bodur}\Dukedesc
\author{A.~Bross}\FNALdesc
\author{E.~Conley}\Dukedesc
\author{Y.~Efremenko}\UTKdesc\ORNLdesc
\author{M.~Febbraro}\ORNLdesc
\author{A.~Galindo-Uribarri}\ORNLdesc\UTKdesc
\author{S.~Gardiner}\FNALdesc
\author{D.~Gonzalez-Diaz}\IGFAEdesc
\author{M.P.~Green}\TUNLdesc\ORNLdesc\NCSUdesc
\author{M.R.~Heath}\ORNLdesc
\author{S.~Hedges}\Dukedesc\TUNLdesc
\author{J.~Liu}\USDdesc
\author{A.~Major}\Dukedesc
\author{D.M.~Markoff}\NCCUdesc\TUNLdesc
\author{J.~Newby}\ORNLdesc
\author{D.S.~Parno}\CMUdesc
\author{D.~Pershey}\Dukedesc
\author{R.~Rapp}\altaffiliation{Now at: Washington \& Jefferson College, Washington, PA, 15301, USA}\CMUdesc
\author{D.J.~Salvat}\IUdesc
\author{K.~Scholberg}\email{kate.scholberg@duke.edu}\Dukedesc
\author{L.~Strigari}\TAMUdesc
\author{B.~Suh}\IUdesc
\author{R.~Tayloe}\IUdesc
\author{Y.-T.~Tsai}\SLACdesc
\author{S.E.~Vahsen}\Hawaiidesc
\author{T.~Wongjirad}\Tuftsdesc
\author{J.~Zettlemoyer}\FNALdesc

\maketitle
\medskip

\medskip

\begin{Abstract}
The Oak Ridge National Laboratory (ORNL) Spallation Neutron Source (SNS) First Target Station (FTS), used by the COHERENT experiment, provides an intense and extremely high-quality source of pulsed stopped-pion neutrinos, with energies up to about 50~MeV. Upgrades to the SNS are planned,  including a Second Target Station (STS),  which will approximately double the expected neutrino flux while maintaining quality similar to the FTS source.  Furthermore, additional space for ten-tonne scale detectors may be available.
We describe here exciting opportunities for neutrino physics, other particle and nuclear physics, and detector development using the FTS and STS neutrino sources.
\end{Abstract}

\snowmass

\def\thefootnote{\fnsymbol{footnote}}
\setcounter{footnote}{0}

\clearpage
\tableofcontents

\clearpage

\section{Introduction}

This white paper is a companion to \textit{The COHERENT Experimental Program}~\cite{coherentwp} Snowmass white paper.  Some material here overlaps with that document; in this document is expanded material specific to the additional physics opportunities that make use of both the First Target Station (FTS) and the Second Target Station (STS) in the era of STS operation.

\section{Neutrinos at the Spallation Neutron Source}
\label{sec:nu-sns}

The Oak Ridge National Laboratory Spallation Neutron Source First Target Station provides neutrons for diverse science goals by colliding GeV protons onto a mercury target.  The protons arrive at the target in pulses several hundred ns wide at 60~Hz.    The proton-Hg collisions create pions; $\pi^-$ are largely captured by nuclei, whereas a very dominant fraction of  $\pi^+$ come to a stop and then decay at rest.  The primary decay products are a monochromatic 30-MeV $\nu_\mu$ and a $\mu^+$ on a short timescale; the $\mu^+$ subsequently decays to a $\nu_e$ and a $\bar{\nu}_\mu$ with well-understood spectra ranging up to $\sim$50~MeV 
(below $\nu_\mu$ charged-current (CC) threshold) with $2.2$-$\mu$s decay time.  The number of neutrinos produced at the FTS amount to approximately 
$5\times 10^{14}$ neutrinos per flavor per MW-s.  The quality of this source is excellent for neutrino physics, given the very high fraction of pions which decay at rest and a pulsed time structure allowing rejection of off-beam backgrounds at the $10^3-10^4$ level.  The COHERENT experiment has already taken advantage of this beam~\cite{COHERENT:2017ipa,COHERENT:2020iec} with detectors deployed 16-25~m from the source in ``Neutrino Alley", an underground corridor parallel to the proton beam, with substantial shielding reducing beam-related neutron flux and 8 meters-water-equivalent overburden.   COHERENT is pursuing multiple physics goals with its suite of detectors~\cite{coherentwp} at the FTS.

\begin{figure}[h!]
    \centering
    \includegraphics[width=0.39\textwidth]{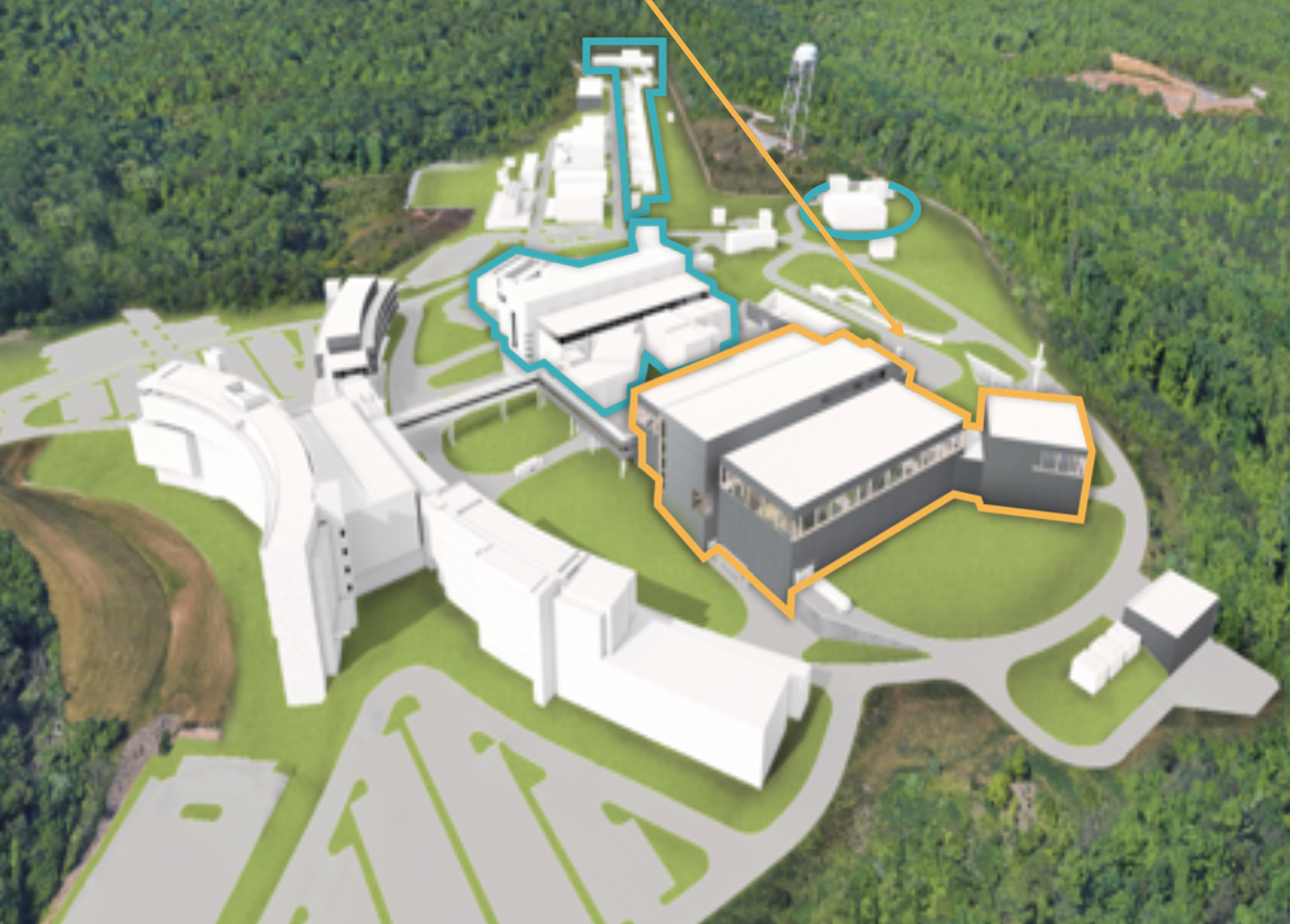}
    \caption{Spallation Neutron Source~\cite{herwig}.  STS buildings are outlined in orange.}
    \label{fig:sns}
\end{figure}

The timing structure of the beam provides not only background rejection, but also opportunities for \textit{flavor separation}.  The prompt $\nu_\mu$ are in-time with the proton beam flux, while the $\bar{\nu}_\mu$ and $\nu_e$ are delayed.  The structure allows well-understood separation of prompt $\nu_\mu$ neutral-current (NC) interactions from delayed $\nu_e$ CC and $\bar{\nu}_\mu$ and $\nu_e$ NC interactions. The timing also gives a handle on systematics for BSM signals for which neutrinos are background (e.g.,~\cite{COHERENT:2021pvd, COHERENT:2022pli}).

\subsection{Planned ORNL Upgrades} 
\label{sec:ornl-upgrades}

ORNL is planning an upgrade to the current 1.4-MW beam.  The Proton Power Upgrade (PPU) project will double the power of the existing accelerator structure, to increase the brightness of pulsed neutron beams and provide new science capabilities.  Furthermore, the Second Target Station (STS) includes a new neutron-production target (of tungsten) along with a new experimental hall and suite of neutron beam lines.  Proton bunches, produced by the same accelerator, will be divided between the FTS and STS in a 3 to 1 ratio. The beam power is expected to be 1.7 MW in 2022 and 2.0 MW in 2024. After STS construction is completed, at the beginning of the 2030's the FTS will receive 2.0 MW at 45 Hz, and the STS will receive 0.7 MW at 15 Hz.  
These upgrades provide exciting new opportunities.  Neutrino flux is approximately proportional to proton power.  Preliminary studies suggest similar per-proton neutrino production rates from the STS as the FTS~\cite{COHERENT:2021yvp}.  Detectors sited in between the STS and FTS, at tens of meter baselines, will receive flux from both.  It will be technically feasible to site 10-tonne-scale detectors at the STS, with sufficient shielding and overburden.

\subsection{The Neutrino Source}
\label{sec:nusource}

\begin{figure}[htbp]\centering
    \includegraphics[width=0.4\textwidth]{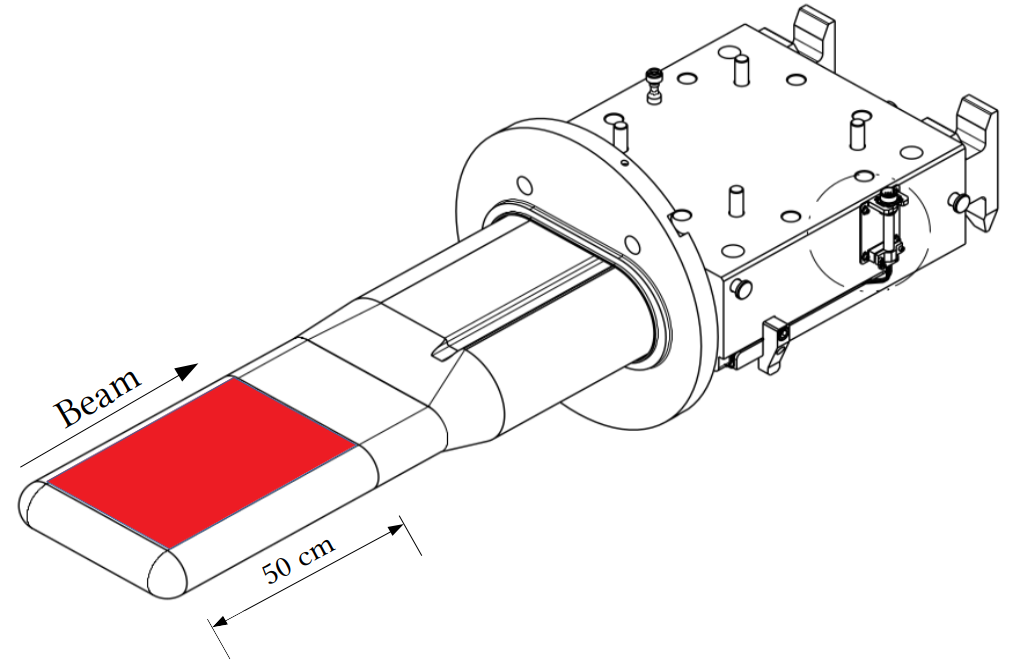}
        \includegraphics[width=0.5\textwidth]{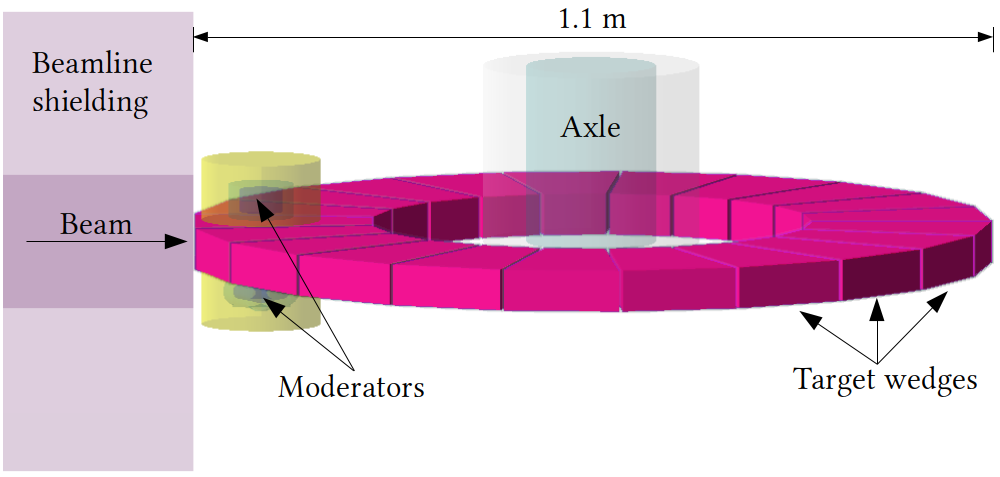}
    \caption{Left: Diagram of the existing Hg target at the FTS; the region modeled in COHERENT's flux simulations is shaded in red. Right: Geant4 model of the preliminary STS design. Figures reproduced from Ref.~\cite{COHERENT:2021yvp}.}
    \label{f:target_diagrams}
\end{figure}

At both the FTS and STS, proton energies will be too low for significant kaon production, so $\pi^+$ decay remains the dominant production source for neutrinos at both target locations. The existing FTS contains liquid mercury in a steel casing with a rectangular interior cross section of $39.9 \times 10.4$~cm$^2$; circulation of the Hg within the target distributes the heat load of the beam. The preliminary STS design takes a different approach to relieving heat load: 21~tungsten wedges are assembled in a ring, 1.1~m in diameter. This ring is rotated between proton deliveries, allowing each wedge to cool before it receives beam again. Fig.~\ref{f:target_diagrams} shows both the FTS and STS targets as modeled in Geant4. 

At a proton energy of 1.3~GeV (as expected following the Proton Power Upgrade), our simulations predict 0.36 $\nu$ per proton delivered to the FTS target, and 0.39 $\nu$ per proton delivered to the STS target, with a 10\% uncertainty driven by our imperfect knowledge of pion-production cross sections at these energies~\cite{COHERENT:2021yvp}. The STS is a slightly more efficient neutrino producer due to the greater density of tungsten compared to mercury. Our FTS simulations show that more than 99\% of all neutrinos are produced from decay-at-rest processes, but this fraction depends on the proton energy and on the details of the beamline and local moderators, since $\pi^+$ are sometimes produced outside the target. Detailed simulations to characterize the STS as a neutrino source will require finalized design information not only for the target structure but also the proton beam window, moderators, and shielding structures.

\begin{figure}[htbp]\centering
    \includegraphics[width=0.45\textwidth]{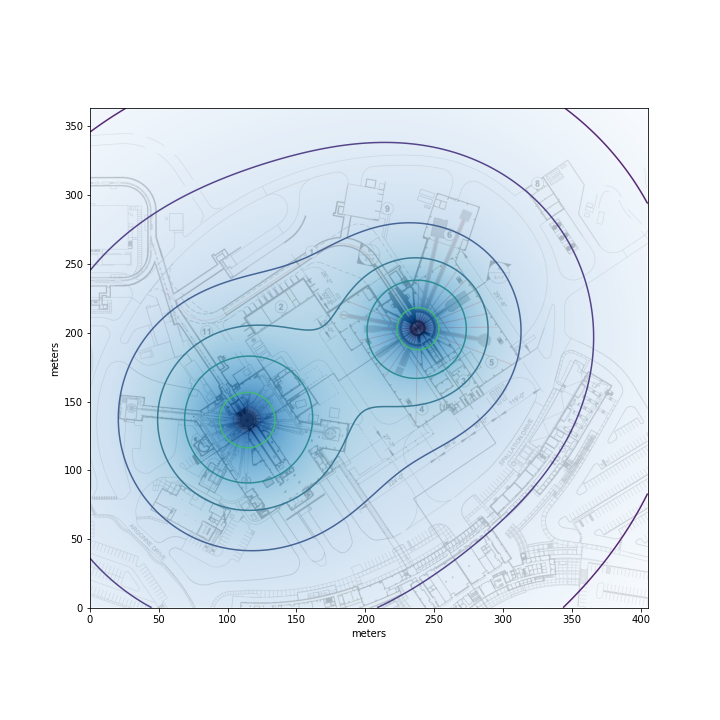}
        \includegraphics[width=0.45\textwidth]{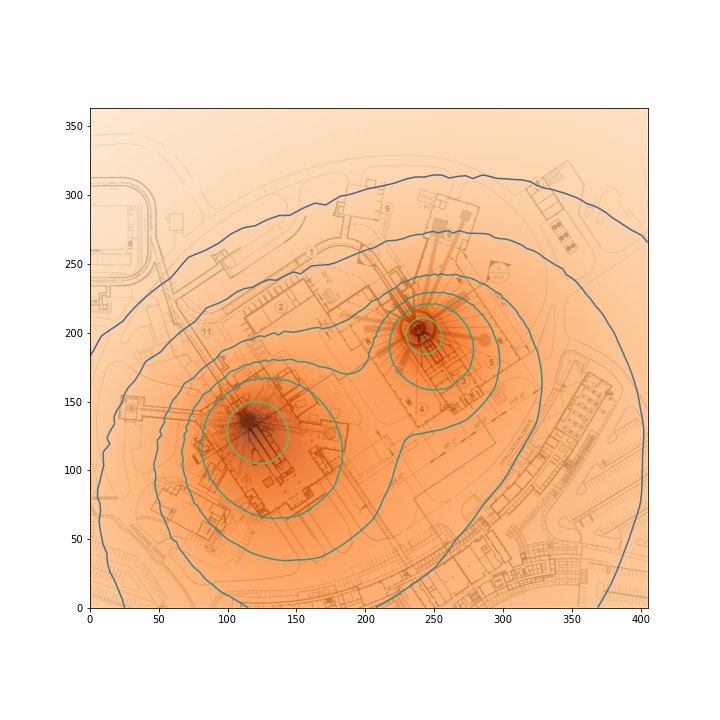}
    \caption{Left: Total neutrino flux emitted from the FTS and STS at 2.4~MW, as a function of position, showing also iso-flux contours. Right:  dark matter flux for 100 MeV mass.  }
    \label{f:fluxes}
\end{figure}

Fig.~\ref{f:fluxes} displays heat maps of the expected neutrino and dark-matter fluxes from the FTS and STS combined; given their close proximity, a single detector can view particles from both sources.

\section{Physics Opportunities}
\label{sec:phys-opportunities}

\subsection{Overview}
\label{sec:overview}

We highlight briefly here multiple motivations for exploitation of the SNS FTS and STS neutrinos~\cite{Bolozdynya:2012xv}. Some of these are described in COHERENT's white paper~\cite{coherentwp}, and so are only touched on briefly here, but others go beyond the scope of COHERENT in Neutrino Alley.

\subsection{Coherent Elastic Neutrino-Nucleus Scattering (CEvNS)}
\label{sec:cevns}

CEvNS is the process in which a neutral-current (NC) neutrino interaction results in the recoil of the nucleus as a whole~\cite{Freedman:1973yd, Kopeliovich:1974mv}.  The COHERENT collaboration's program of CEvNS measurements for a range of nuclear targets, testing the SM-predicted $N^2$ dependence of the cross section (where $N$ is the neutron number of the nucleus), has potential for a wide range of physics~\cite{Abdullah:2022zue}. CEvNS is a sensitive probe of non-standard interactions (NSI) of neutrinos with heavy and light mediators. It can provide measurements of $\sin^2\theta_W$ and of neutrino electromagnetic properties.  A percent-level precision, CEvNS can probe nuclear structure with unprecedented sensitivity.  Furthermore,  CEvNS is an effective tool for sterile neutrino oscillation searches~\cite{Garvey:2005pn,  Anderson:2012pn}, as discussed in Sec.~\ref{sec:sterilenu}.

\begin{figure}[!tb]\centering
  \includegraphics[width=0.9\linewidth]{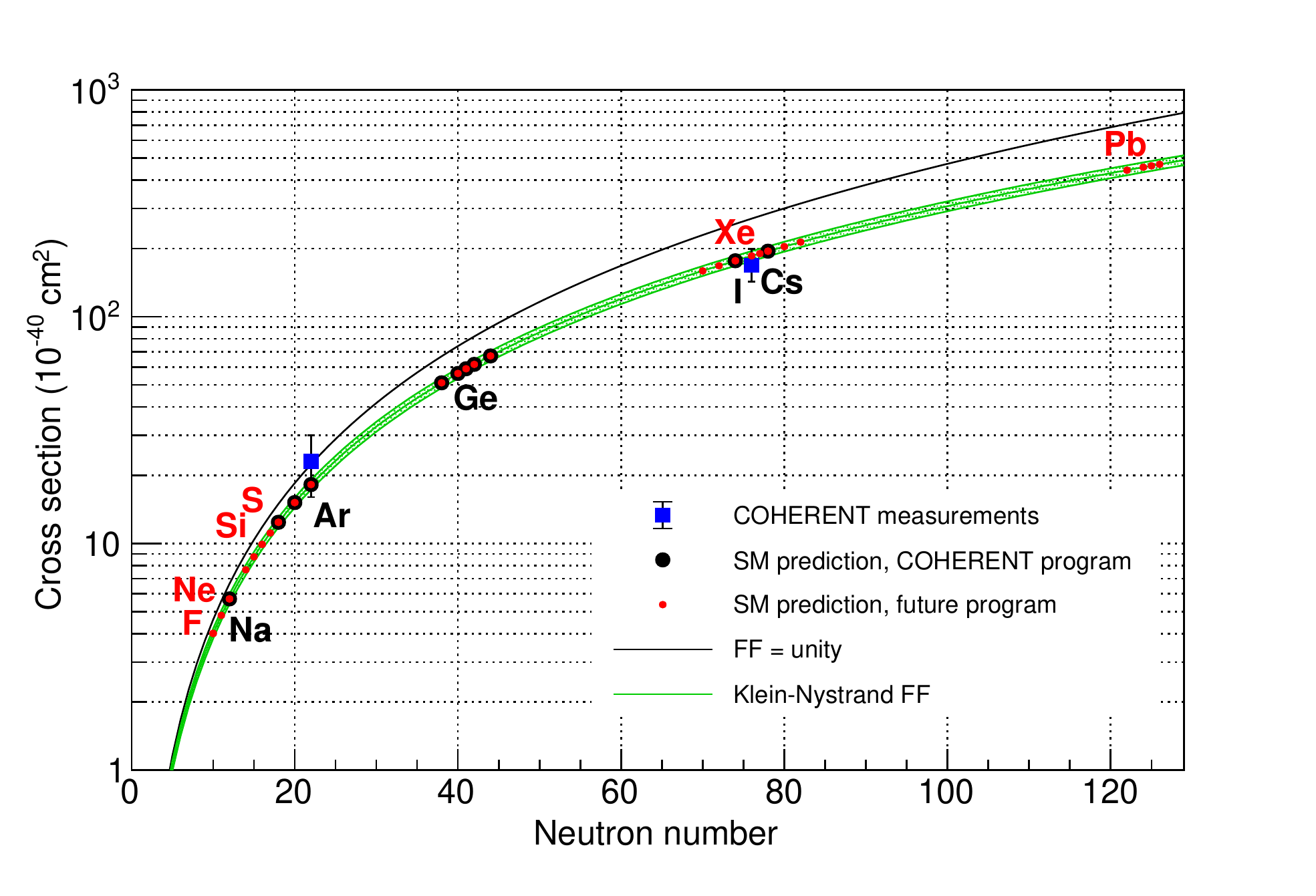}
  \caption{Flux-averaged CEvNS cross sections as a function of neutron number, from Ref.~\cite{Barbeau:2021exu}, with several potential future nuclear targets highlighted in red.  The highlighted targets are not intended to be a comprehensive set, but rather several for which particular, known-to-be-feasible detector technologies have been proposed for CEvNS measurements. }
  \label{fig:targets}
\end{figure}

More details of specific physics studies are given in Ref.~\cite{coherentwp}, but we show in Fig.~\ref{fig:targets} some examples of possible future nuclear targets over a range of $N$ values.

\subsection{Accelerator-Produced Dark Matter}
\label{sec:accelDM}

In addition to neutrino-induced CEvNS recoils, low-threshold detectors at the STS could also observe nuclear recoils induced by any light dark matter particles that may be produced in the target.  SNS experiments are sensitive to dark matter masses below the SNS beam energy, which will be 1.3~GeV during STS running.  Thus, any dark matter particles observed would be below the Lee-Weinberg bound, necessitating an additional mediator particle between standard model and dark matter particles.  CEvNS experiments can ambitiously probe dark matter particles mediated by a vector that kinetically mixes with the photon~ \cite{PhysRevD.70.023514,BOEHM2004219,Pospelov:2007mp,PhysRevD.84.075020} or hadronically~\cite{Aranda:1998fr,Gondolo2012LightDM,Batell:2014yra,deNiverville:2015mwa,deNiverville:2016rqh,Boyarsky:2021moj}.  The first result from the COHERENT 14.6~kg CsI[Na] detector has already placed the most stringent bounds on dark matter in the 10-100~MeV region~\cite{COHERENT:2021pvd}.  Taking advantage of detector halls designed specifically for CEvNS experiments at larger mass scales and with better neutron shielding at the STS can significantly improve reach.

Mediator particles, $V$, would be dominantly produced through meson decay in flight at the SNS through $\pi^0\rightarrow\gamma V$ and $\eta^0\rightarrow\gamma V$ with dark matter particles, $\chi$, produced through $V\rightarrow\bar{\chi}\chi$ decay.  As these decays occur in flight, the dark matter flux would be prompt, coincident with the $\nu_\mu$ flux.  As such, using timing, the CEvNS background can be mitigated.  Further, the delayed timing region can be used to constrain uncertainties on the neutrino interaction model and detector response to low-energy nuclear recoils.  This ensures that dark matter searches are limited by statistical uncertainty with detectors even 1000$\times$ the COHERENT CsI[Na] detector.  Since the dark matter is produced from decay in flight, it is boosted in the forward direction while the neutrino flux is isotropic, as illustrated in Fig.~\ref{f:fluxes}.  The degree of directionality is correlated with the dark matter particle mass.  Thus, the dark matter signal to CEvNS background is highest in the forward direction and, for optimal dark matter sensitivity, a detector should be built within about 20$^\circ$ of the beam direction.  This alone gives a significant sensitivity gain compared to that accessible in Neutrino Alley at the FTS.  Additionally, with multiple detectors implemented, the nature of a dark matter excess can be tested by measuring the dark matter to CEvNS ratio at different off-axis angles.

\begin{figure}[!tb]\centering
  \includegraphics[width=0.8\linewidth]{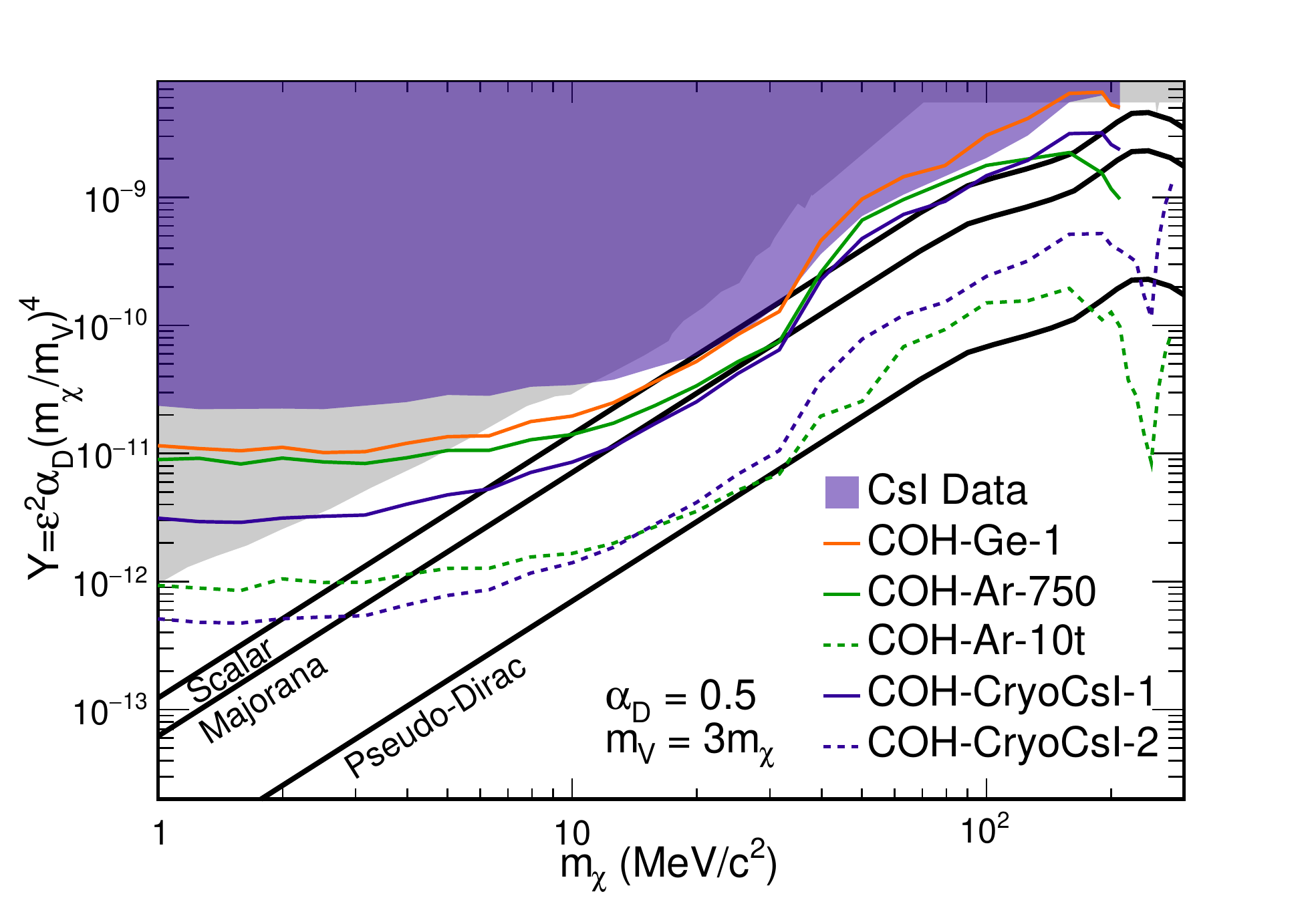}
  \caption{Sensitivity to discover vector-portal dark matter at the SNS.  The dashed curves give the sensitivity to the proposed argon calorimeter and cryogenic CsI scintillator.  For comparison, the reach of COHERENT at the FTS is shown by the solid curves.  The shaded purple and grey regions give current constraints from COHERENT and other experiments, respectively.}
  \label{f:DarkMatter}
\end{figure}

We estimated the sensitivity of two CEvNS detectors for discovering dark matter particles produced at the SNS through detection of their induced nuclear recoils.  The most sensitive detectors we evaluated are the 10-tonne argon scintillation calorimeter and the 700-kg cryogenic CsI scintillator.  The argon detector is assumed to be sited 20 m from the beam target at an off-axis angle of 20$^\circ$.  The CsI detector is placed orthogonal to the beam.  With two significantly different off-axis angles, these two detectors will test the angular dependence of the dark matter flux in the event of a positive signature.  The CsI detector has a significantly lower nuclear recoil threshold giving this detector better sensitivity to low-mass dark matter.  The argon detector is more beneficial for high-mass dark matter due to increased mass and is thus chosen for the on-axis position.  As the argon detector is also designed to allow measurement of neutrino inelastic interactions with several-MeV or more visibile energy, we may also be able to discover dark matter by identifying inelastic interactions with argon nuclei by isolating nuclear deexcitation gamma rays, though work is preliminary~\cite{Dutta:2022tav}.  We estimate sensitivity of these detectors by performing a 2D log-likelihood fit in recoil energy and recoil time.  All standard-model backgrounds, dominated by beam-unrelated backgrounds and CEvNS, are included in the fit.  Though there is no signal predicted in the delayed time region, these events are included in the fit to constrain background uncertainties relevant for the dark matter ROI.  The predicted sensitivity to constraining dark matter at the STS is shown in Fig.~\ref{f:DarkMatter}.

New facilities available at the STS will allow aggressive tests of sub-GeV accelerator-produced dark matter at the SNS.  Fig.~\ref{f:DarkMatter} shows the constraint assuming $\alpha_D=0.5$.  This is a conservative choice as higher values become non-perturbative and lower values yield stronger constraints relative to the cosmological expectations~\cite{deNiverville:2015mwa}.  The couplings consistent with the dark matter relic abundance are shown as black lines.  These expectations depend on the spin and mass phenomenology of the dark matter particle.  At the STS, expected sensitivity lies near the most conservative scenario, a pseudo-Dirac fermion yielding powerful discovery potential, particularly for dark matter masses between 10 and 300 MeV/c$^2$.

\subsection{Axions}
\label{sec:axionsBSM}

The STS can be used to search for axion-like particles (ALPs) by utilizing their large photon flux. Traditional searches for pseudoscalar ALPs rely on their decay in beam dumps or their conversion into photons in haloscopes and helioscopes. Through Primakoff-like or Compton-like channels at stopped pion experiments, the sensitivity to the ALP-photon/electron couplings can improve upon existing limits~\cite{Batell:2022xau}. 

\subsection{Inelastic Neutrino-Nucleus Cross Sections}
\label{sec:inelastics}

Neutrinos from stopped pions overlap significantly with the expected energy range of neutrinos from  core-collapse supernovae, which go up to several tens of MeV.  The SNS therefore offers excellent opportunities for the study of neutrino-nucleus interactions of relevance for supernova neutrino detection, as well as for understanding of processes within the supernova itself, including both astrophysical mechanisms and nucleosynthesis.   The energy regime of interaction products is of the same order as the neutrino energies.  Target nuclei of particular interest (i.e., relevant for existing or planned supernova neutrino detectors, as well as for understanding of nucleosynthesis)  are Ar, O, Pb, Fe and C.  

Solar neutrinos, which have energies up to about 15~MeV, are an interesting physics target for DUNE~\cite{Capozzi:2018dat}, and knowledge of the CC $\nu_e^{40}$Ar cross section will enable interpretation of that signal also.

Measurement of inelastic neutrino-nucleus cross sections is also of intrinsic interest for study of the weak interaction and for nuclear structure physics~\cite{coherentwp}.

A specific example of particular relevance to the U.S. community is argon, the material to be used in DUNE. DUNE has excellent sensitivity to the $\nu_e$ component of a supernova neutrino burst~\cite{Abi:2021vq}, for which an observation will yield rich physics and astrophysics.  However, there currently exist \textit{no} measurements of neutrino cross sections on $^{40}$Ar in the relevant energy range.  The dominant $\nu_e$ CC as well as the very-poorly-understood NC excitation cross sections are both of great interest.  Uncertainties in these cross sections limit the quality of information which can be extracted from a burst observation.
 
Some existing and near-future COHERENT detectors, while optimized for low-energy recoils, have sufficient dynamic range to study some of these processes in argon, oxygen, lead and iron.  However for precision cross section measurements, full understanding of the distribution of final-state interaction products, fine-grained tracking detectors will be needed. 

\subsection{Sterile Neutrinos}
\label{sec:sterilenu}

The large neutrino flux produced from $\pi^+$ decay-at-rest at the SNS is well suited for testing the LSND excess~\cite{LSND:2001aii} which could be explained by an additional sterile neutrino state beyond the three-flavor paradigm.  MiniBooNE later saw similar excess~\cite{MiniBooNE:2020pnu} while observation of neutrino disappearance at nuclear reactors~\cite{Mention:2011rk} and in gallium experiments~\cite{SAGE:1998fvr,Abdurashitov:2005tb,Kaether:2010ag} are also consistent with the sterile neutrino hypothesis.  In the context of sterile neutrino oscillations, detector baselines chosen for CEvNS experiments at the STS lie at the first oscillation maximum for $\Delta m^2_{41}\approx2$~eV$^2$ for the monoenergetic $\nu_\mu$ flux component of the SNS flux, consistent with current global fits of all oscillation data~\cite{Gariazzo:2017fdh}.  As such, a CEvNS detector deployed at the STS can test the sterile neutrino explanation of the LSND anomaly.  

Generally, CEvNS experiments are favorable tests of the sterile neutrino hypothesis with several key advantages.  The CEvNS cross section is well understood theoretically allowing precision measurements of NC disappearance of both $\nu_e$ and $\nu_\mu$ flavors.  Oscillation effects for each flavor can be isolated using timing information of CEvNS recoils allowing sensitivity to both $\theta_{14}$ and $\theta_{24}$ within the same experiment.  Further, the prompt $\nu_\mu$ flux is monoenergetic, giving an efficient test of the mass mixing parameter $\Delta m^2_{41}$.  Though CEvNS is a NC process, the observable recoil energy is loosely correlated with neutrino energy as the maximum recoil energy is $\sim 2E_\nu^2/M_\text{nuc}$.  Thus, the highest energy recoils originate from a narrow-band neutrino flux with a cutoff of $m_\mu/2=52.8$~MeV.  The dominant uncertainty on predictions of the CEvNS rates is the neutrino flux normalization.  This uncertainty can be mitigated by studying the $L/E$ dependence of the oscillation phase 
with multiple detectors on different baselines, a strategy currently employed at the FTS.  With concurrent operation of both targets at the SNS, the $L/E$ dependence can instead be tested by considering data from both targets separately using the same detector, and mitigating detector response uncertainties.  

\begin{figure}[!tb]\centering
  \includegraphics[width=0.8\linewidth]{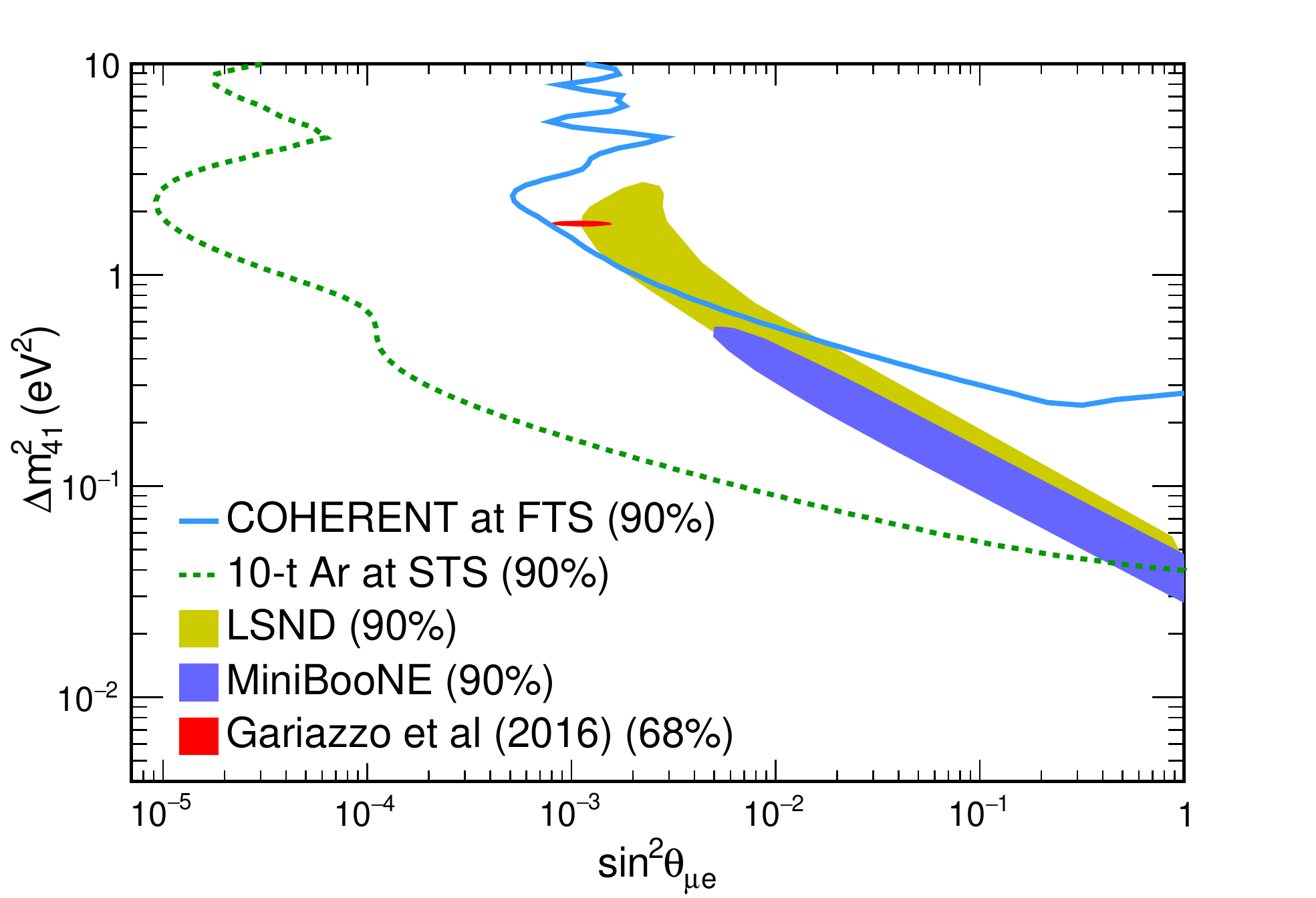}
  \caption{Sensitivity to sterile neutrinos for a 10-tonne argon scintillating calorimeter at the STS compared to the global preferred sterile neutrino parameter space along with individual results from LSND and MiniBooNE.  The reach is very favorable with potential for CEvNS measurements to probe mixing angles two orders of magnitude lower than current best fits.  The reach of COHERENT at the FTS is also shown.}
  \label{f:SterileSens}
\end{figure}

A 10-tonne liquid argon scintillating calorimeter placed 20~m from the STS and 120~m from the FTS would be sufficient to record roughly 1000 CEvNS events from neutrinos produced at the FTS target per year, high enough to probe the scattering rate to $2\%$ after five years of data.  We show the expected sensitivity of this detector after five years in Fig.~\ref{f:SterileSens} when fitting in time, to isolate different neutrino flavors, and argon recoil energy, correlated with incident neutrino energy.  By significantly reducing systematic uncertainties, CEvNS experiments can search for $\approx1\%$ disappearance of either the $\nu_\mu$ or $\nu_e$ flux in the SNS flux, sufficient for probing two orders of magnitude beyond the global best fit of $\sin^22\theta_{\mu e}=\sin^22\theta_{14}\sin^2\theta_{24}$.  

\section{Possible Detectors}
\label{sec:possibledets}

Possible new detectors include those sensitive to keV-scale recoils at the tens of kg to 10-tonne scale.  These include scale-ups or upgrades of those deployed or proposed for the FTS by COHERENT~\cite{coherentwp}, such as cryogenic crystals, germanium, sodium-iodide, or large noble-liquid detectors, as well as large single-phase or dual-phase noble liquid time projection chambers (TPCs).  Such detectors may also be sensitive to inelastic interactions~\cite{coherentwp} as well.   A large heavy-water detector for flux normalization~\cite{COHERENT:2021xhx}, similar to that planned for COHERENT in Neutrino Alley, is another possibility, which would also have sensitivity to inelastic interactions on $^{16}$O.  Different types of gaseous TPCs are additional interesting possibilities.

Other possible detector materials and configurations include liquid scintillator, lead-, iron- or copper-based detectors, silicon-based detectors, directional CEvNS detectors, scintillating bubble detectors, and low-threshold bolometers.  Generally, good event-by-event timing resolution is desirable, in order to take advantage of the background rejection afforded by the beam timing.

We describe in the following sections a handful of specific concepts, but recognize that there are broader possibilities for deployment at the STS.

\subsection{Single-Phase Argon}
\label{sec:ar-1phase}

\begin{figure}[hptb!]
\centering
\includegraphics[width=0.8\textwidth]{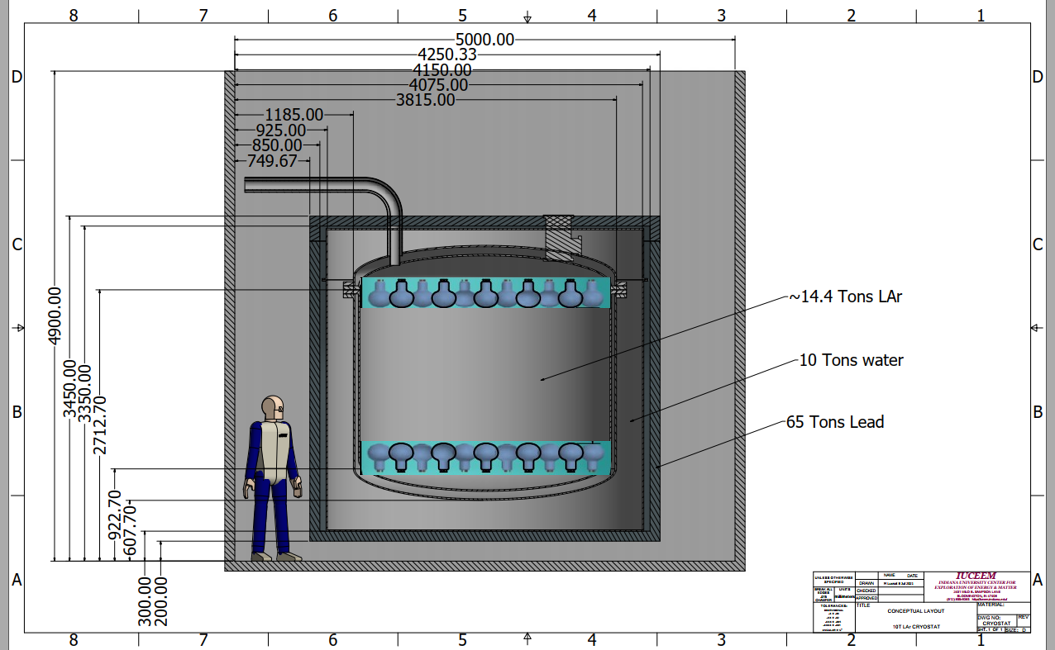}
\caption{A dimensioned conceptual design of a shielded $\approx10$~ton LAr scintillation detector sited within a room at the SNS STS.}
\label{fig:10tonLArconcept}
\end{figure}

An $\approx10$-tonne single-phase liquid argon detector in a well-shielded location, within $\approx20$~m of the $\approx1$~MW neutrino source will enable an impressive suite of high-sensitivity, low-energy physics measurements such as the study of NSI with CEvNS,  accelerator-produced dark matter, and sterile neutrino searches, as described in Sec.~\ref{sec:phys-opportunities}. A conceptual design for such a detector is shown in Figure~\ref{fig:10tonLArconcept}. This particular configuration shows the single-phase liquid scintillation detector with 8" PMT readout. One may also consider using SiPMs for light collection in full, or sparingly. to provide additional detectors to increase the detector dynamic range and perhaps allow for better direction reconstruction for inelastic events with Cherenkov light.   
The same cryostat configuration, and to first approximation, the same support utilities, would allow a LAr TPC as well.

The space of approximately $5 \times 5 \times 5$~m$^3$ would accommodate this detector but would require assembly in another location.  This is required to maintain our strict shielding requirements to allow a low-energy experiment near such a high-power stopped pion neutrino source. The COHERENT experiment~\cite{COHERENT:2018gft}, currently running at the SNS FTS, has made measurements of signals and backgrounds.  This experience will enable optimized plans for large detectors at the SNS within a dedicated facility at the SNS STS.

In addition to the required shielding for a low-background location near the STS, many aspects of the detector design should be optimized for maximum physics sensitivity.  For LAr, an important consideration is the detector medium itself. Argon sourced in the usual manner derives from the atmosphere and contains $\approx 1$~Bq/kg of $^{39}$Ar, a $\beta$-emitter which contributes to the steady-state background underlying the beam-related signal. A significant amount of work has gone into finding a source of underground argon~\cite{Alexander:2019uvv}, depleted in $^{39}$Ar, which would significantly improve the sensitivity of a large LAr detector as considered here and should be used if possible.  More improvement in LAr light collection is possible via the technique of xenon doping~\cite{Akimov:2019eae} and more advanced photodectors such as SiPMs.

\subsection{LArTPCs}
\label{sec:lartpc}

A single-phase liquid argon time projection chamber (LArTPC) deployed at the SNS will provide an ideal detector capable of precision neutrino and neutron measurements relevant to the low energy physics regime of the future planned Deep Underground Neutrino Experiment (DUNE). An $\mathcal{O}$(tonne)-scale detector with sufficiently low threshold would be capable of making cross-section measurements of low energy charged-current (CC) electron neutrino ($\nu_e$) interactions ($\nu_e + {}^{40}{\rm Ar} \rightarrow e^- + {}^{40}{\rm K^*}$), possibly a relatively low statistics measurement of electron neutrino--electron (${\nu_e-e^{-}}$) elastic scattering, as well as searching for other inelastic (non-CEvNS) NC interactions.  As DUNE's far detector is based on the technology of LArTPCs, measurements with the same technology can be directly translated as relevant input to the supernova,  solar, and atmospheric neutrino analyses planned to take place for DUNE~\cite{Caratelli:2022llt}.

In a TPC, as charged particles traverse the bulk material, they produce ionization electrons and scintillation photons. An external electric field allows the ionization electrons to drift towards the anode of the detector and be collected on a charge-sensitive readout. The charge typically moves over meters of distance on the time scale of milliseconds (with a nominal drift velocity at 500 V/cm of 1.6 mm/$\mu$s.) The combined measurement of the scintillation light, providing the interaction time $t_0$, and the arrival time of the ionization charge, allows for the 3D reconstruction of the original charged-particle topology within the TPC. The light signal forms from the de-excitation of the argon on time scales of nanoseconds to microseconds. Thus the TPC provides a fully active tracking detector with calorimetric reconstruction capabilities in the absence of instrumention of the bulk volume of the detector.

Conventional methods for reading out the ionization charge in a LArTPC rely on the use of consecutive planes of sensing wires, with the 2D projections providing two of the three spatial coordinates needed to reconstruct the 3D image. This method has been deployed in many recent LArTPC experiments \cite{Anderson:2012vc,LArIAT:2019kzd,Bian:2015qka,antonello2015operation,acciarri2017design}. While this method of readout has been successful in several detectors, it suffers from an intrinsic limitation in resolving ambiguities for various topologies. Novel event reconstruction techniques may be employed to overcome these difficulties~\cite{Qian:2018qbv, MicroBooNE:2020vry, MicroBooNE:2020jgj, MicroBooNE:2021zul, MicroBooNE:2021ojx} but are often very complex and not readily adaptable to diverse experimental setups. Another intrinsic limitation of the wire-based readout is that the use of long sense wires introduces significant capacitance to the readout electronics~\cite{MicroBooNE:2017qiu}, which may limit the extraction of physics signals at low-energy thresholds [e.g., $\mathcal{O}(\leq$MeV)]~\cite{LArIAT:2019gdz}. 

A pixel-based readout scheme had not been previously considered for large LArTPCs because of the increased number of readout channels, and the high data rate and power consumption.  A transformative step realized by the LArPix \cite{Dwyer:2018phu} and Q-Pix \cite{Nygren:2018rbl} consortia now allows one to build a fully pixelated low-power charge readout capable of efficiently and accurately capturing signal information. The use of a 3D-based pixel readouts can offer significant improvements~\cite{Adams:2019uqx} in the reconstruction of events in a LArTPC and has been shown to offer enhancements for low energy neutrino physics \cite{Q-Pix:2022zjm} as well as the ability to operate and reconstruct in various prototypes \cite{roberto_mandujano_2022_6805002}. The deployment of a $\mathcal{O}$(tonne)-scale pixel-based LArTPC detector can allow for unique measurments at ORNL's SNS. It could also serve as a test-bed for new LArTPC technology, such as novel pixelated readouts and novel photodetection schemes to be realized in a pixel-based detector.

\begin{figure}
    \centering
    \includegraphics[scale=0.28]{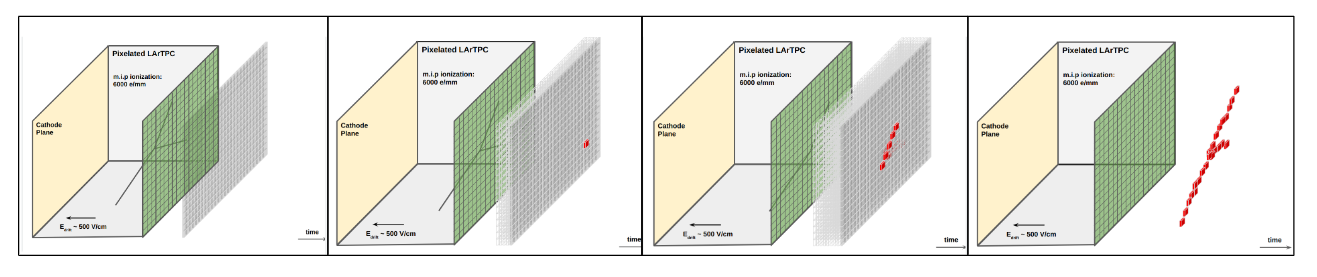}
    \caption{Illustration of the intrinsic 3D readout enabled by a pixelated LArTPC. Each frame represents a snapshot in time of the charge arriving on the individual pixels (shown in red). The fully 3D position of the charge depositions is reconstructed with a spatial resolution dictated by the pixel pitch and charge arrival time resolution.}
    \label{fig:pixel3d}
\end{figure}

\begin{figure}
    \centering
    \includegraphics[scale=0.3]{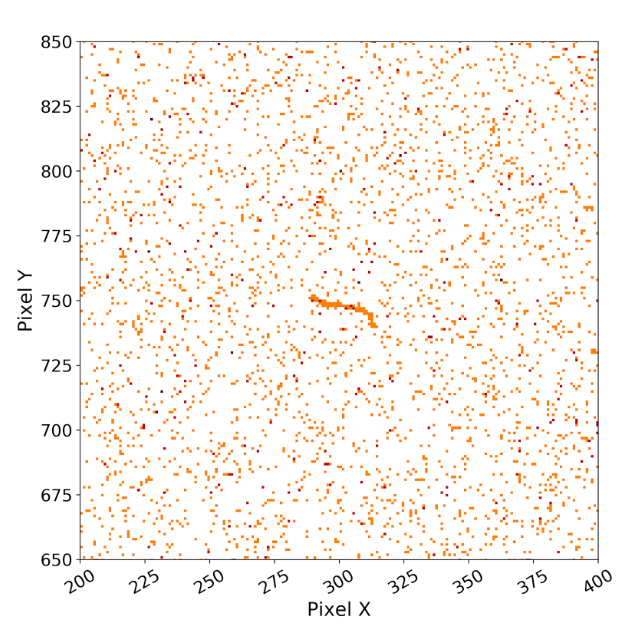}
    \caption{Example visualization of a pixelated LArTPC plane readout showing a 30-MeV electron track, including radiological backgrounds.}
    \label{fig:lartpclowenergynu}
\end{figure}

Inelastic interactions of MeV-scale neutrinos leave only short
energy depositions surrounded by a number of small depositions of charge (often referred to as ``blips''). These blips can come from bremsstrahlung photons later interacting by Compton scattering, MeV-scale photons  emitted from inelastic scattering of neutrons
some distance away, de-excitation of argon nuclei, charge via Compton scattering~\cite{PhysRevD.99.012002,1748-0221-12-09-P09014}, as well as any intrinsic radiological backgrounds. A simulation of a 30=MeV electron neutrino interaction in the presence of radiological backgrounds is shown in Fig.~\ref{fig:lartpclowenergynu}. For the purposes of this illustration, 10 seconds of charge information have been compressed into a 2D representation of the data where the short energy deposition due to the electron can clearly be seen as well as a number of smaller ``blips'' produced by the various backgrounds.

A LArTPC designed with pixelated charge collection system has an advantage
for tackling the challenges of reconstructing short tracks and blips of charge
in a high-multiplicity environment where sizable cosmic ray and beam-induced
background occurs\cite{instruments5040031}.
The pixelated charge collection system natively provides a 3D event topology,
while the scintillation light signals from the primary interaction can allow for separation of the interaction from the other backgrounds.

As the blips of charge may be shorter than the pitch of the charge collection system,
the amount of deposited charge in a pixel can often be lower than
that from a long, minimal-ionization particle.
It is important to take the feature into account in design of the charge collection
system.
For example, the requirements on the signal-to-noise ratio need to be evaluated
accordingly, allowing the charge deposition from physics-signal activities to be
distinguished from noise.
The threshold and dynamic range of the pixel readout may also have to be reconsidered
reflecting the need to select such signals.
Calibration and resolution studies will also have to be specifically developed for such
MeV-scale charge deposition.
The development of pixelated charge collection systems will be in synergy with
the ongoing efforts of the DUNE LArTPC near detector and the proposed GammaTPC
which aims to detect MeV-scale $\gamma$ rays in space~\cite{https://doi.org/10.48550/arxiv.2203.06894}.
Specifically, GeV-scale neutrinos, the neutrinos produced by accelerators
in DUNE, leave a number of MeV-scale charge deposition from the electromagnetic
showers and from neutron scattering~\cite{PhysRevD.99.036009},
R\&D and measurements targeting MeV-scale charge deposition
will allow elaborate studies,
which will further improve the energy reconstruction of oscillation
measurements in DUNE.

The opaque pixelated charge collection system makes it impossible
to keep the conventional TPC design,
in which the light detectors are mounted behind the wire-based charge readout planes.
A few R\&D projects are currently progressing, pursuing a few percent of
light detection efficiency and a few nanoseconds of time resolution~\cite{Rooks:2022vrl,instruments5010004,instruments2010003,Anfimov_2020}.

LArTPCs located at surface,
such as the experiments in the Short Baseline Neutrino (SBN) Program,
suffers from the amount of cosmic rays traversing
its volume in the millisecond time scale.
Hence, they are usually accompanied by scintillator panels and sometimes by overburden
to, respectively, identify cosmic muons and mitigate secondary particles
induced by cosmic rays~\cite{MicroBooNE:2019lta,Machado:2019oxb}.

\subsection{Heavy Water}

The neutrino-flux normalization, with its 10\% uncertainty, is currently the leading systematic for COHERENT's CEvNS measurements at the FTS~\cite{COHERENT:2020iec, Akimov:2021dab}. This systematic affects all neutrino-interaction cross-section measurements across all detector types. Geant4 simulations give insight into the behavior of the neutrino flux at different detector locations or with different beamline elements or beam properties, but cannot by themselves reduce this 10\% uncertainty, which arises from uncertainty in the pion-production cross section for protons in the 1-GeV energy range~\cite{COHERENT:2021yvp}. There are two possible approaches for reducing this uncertainty: thin-target pion-production measurements at dedicated beamlines, and direct benchmarking of the neutrino flux using a known interaction. Both approaches are important, but thin-target measurements cannot adequately address this uncertainty by themselves, because incident protons can still produce $\pi^+$ and hence neutrinos even after losing significant energy in the dense Hg and W targets. Neutrino-flux calculations for the FTS and STS must convolve the pion-production cross section with the energy-loss profile of protons in the target. Pion-production cross sections can, in principle, be measured at a range of energies below the nominal SNS beam energy, but it is infeasible to access the proton energy-loss profile.

In order to benchmark the neutrino flux at the FTS during and after the proton power upgrade, COHERENT is constructing a two-module heavy-water detector~\cite{COHERENT:2021xhx}. Each module functions as a water Cherenkov detector, detecting light from the high-energy electron produced by the charged-current interaction
\begin{equation}
    \nu_e + d \rightarrow p + p + e^-
\end{equation}
This interaction is understood theoretically at the 2--3\% level~\cite{Mosconi:2007tz, Ando:2019yum}, with active ongoing effort to reduce the uncertainty. 

Given the space constraints of Neutrino Alley at the FTS, the heavy-water modules are constructed as nested, upright cylinders. D$_2$O is contained inside a transparent acrylic cylinder, surrounded by a 10-cm H$_2$O tail catcher that is in turn contained within a steel vessel. The inner surfaces of the vessel are coated in highly reflective Tyvek, and Cherenkov light is viewed from above by 12 PMTs. This limited photocathode coverage sacrifices some energy resolution and fiducialization ability, but is efficient in terms of cost and space.

\begin{figure}[htb]
\label{fig:d2o-sensitivity}
\centering
\includegraphics[width=0.6\textwidth]{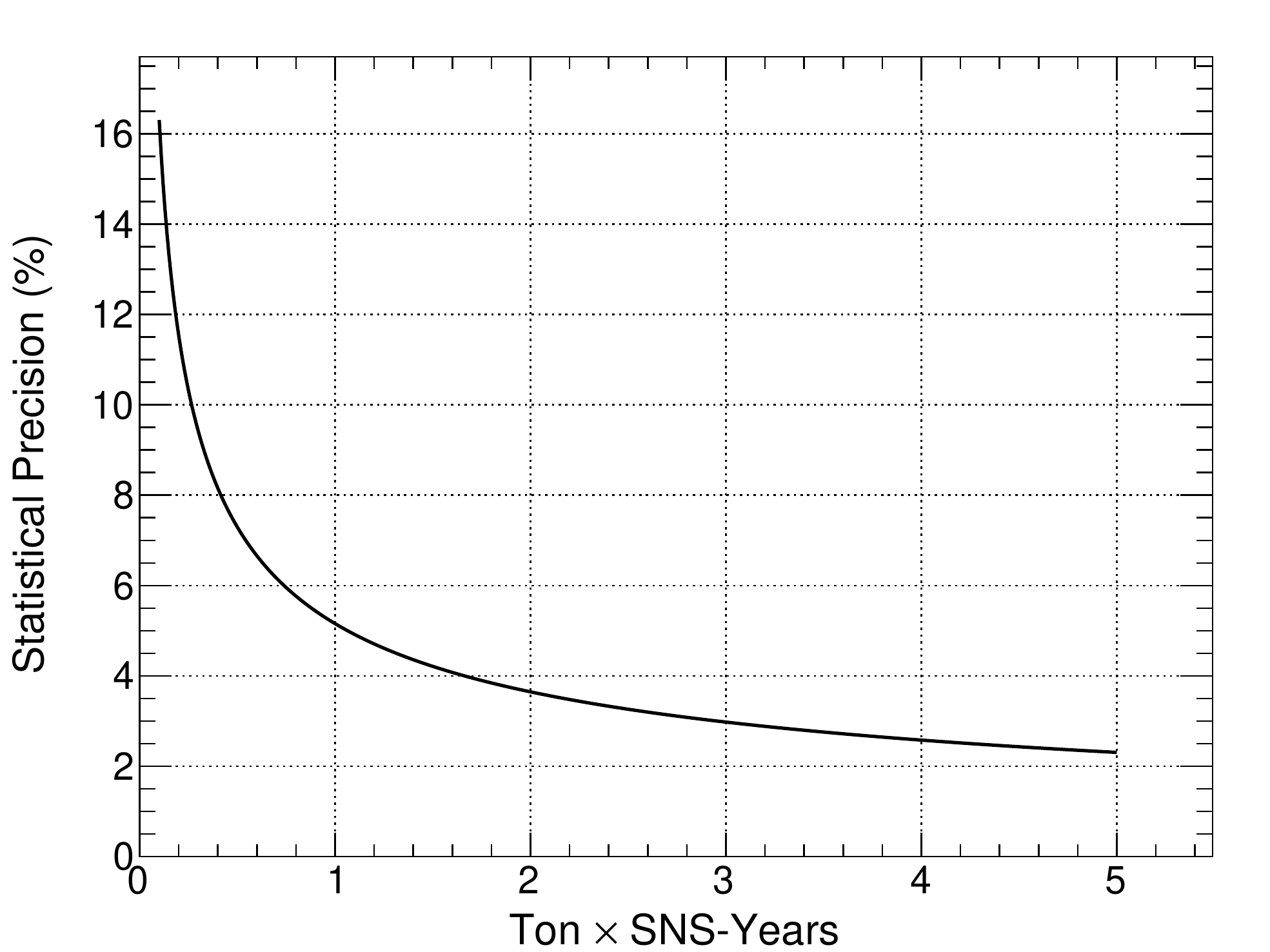}
\caption{Projected statistical sensitivity over time for the D$_2$O  detector at the FTS, assuming 5000 hours per year of 1.4~MW operations. Reproduced from Ref.~\cite{COHERENT:2021xhx}.}
\end{figure}

Depending on space availability and usage at the new facility, neutrino-flux normalization at the STS could be achieved by simply relocating the two-module D$_2$O detector from the FTS; by this time, the detector will be thoroughly understood. If additional space is available, more precise flux benchmarking can be achieved via a redesign: increasing the D$_2$O volume would improve statistics while increasing the photocathode coverage would improve energy resolution, allowing better separation of the deuterium signal from the dominant background (charged-current interactions on oxygen). As a benchmark, Fig.~\ref{fig:d2o-sensitivity} shows the projected statistical sensitivity for the two-module detector at the FTS, as a function of exposure (ton-SNS-years). Note that the ``SNS-year'' measurement assumes the 1.4~MW operations typical before the proton power upgrade. Despite this upgrade, the STS will run at a lower effective power of about 0.7~MW, since it will receive only one in every four proton pulses. In order to achieve the same statistical sensitivity with the same amount of D$_2$O, the required time in calendar years is nearly doubled\footnote{The relationship is not exact because the tungsten STS is expected to produce about 8\% more neutrinos per proton on target than does the mercury FTS (Sec.~\ref{sec:nusource}).}. The greater statistical challenges of the STS environment increase the appeal of a larger D$_2$O detector, better optimized for precise measurements.

An upgraded D$_2$O detector will also target an improved measurement of the inelastic CC oxygen interaction $\nu_e + {}^{16}\mathrm{O} \rightarrow e^- + {}^{16}\mathrm{F}^*$ (Sec.~\ref{sec:inelastics}), as well as potentially NC excitations,  which are of interest for supernova detection in large water Cherenkov detectors~\cite{Super-Kamiokande:2017yvm, Hyper-Kamiokande:2018ofw}.

\subsection{A Magnetized Gaseous TPC for Low-Energy Neutrino Scattering}
\label{sec:magnetizedGasTPC}

Gas TPCs offer many attractive features for the study of low-energy neutrino scattering, both for coherent and inelastic scattering.  In particular, the target nucleus (main gas component) can be varied from He to Xe, operation at pressures from 0.1 Bar to 10 Bar are readily achievable (potentially within the same system) and the addition of a magnet field allows for spectrometry.  In addition the possibility exists for actual tracking of the nuclear recoil (target nucleus dependent), although this would likely preclude high-pressure operation and thus limited statistics would be available.

\begin{figure}[h]
\label{fig:event}
\centering
\includegraphics[width=0.6\textwidth]{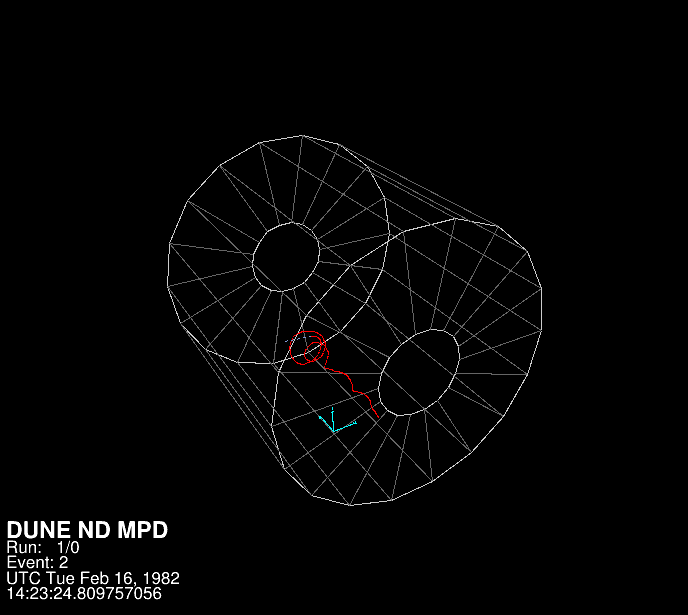}
\caption{A 30 MeV $\nu_e$ CC interaction in a 5-bar gaseous Ar TPC in a 0.1T magnetic field.  The final-state electron shown in the figure has an energy of 15 MeV.}
\end{figure}

Regarding spectrometry, $\nu_e$CC events with E$_\nu$ = 30 MeV have been simulated using DUNE's GArSoft application.  The high-pressure gas TPC (HPgTPC) near detector for DUNE (5m diameter $\times$ 5m long) was used for this, but the pressure was reduced to 5 bar and the magnetic field to 1000G.  An image of one of the events is shown in Figure~\ref{fig:event} where the final-state electron (15 MeV) is clearly visible.
A R\&D program for DUNE's HPgTPC is ongoing and we feel that a parallel but coordinated effort to develop a gas TPC concept for low-energy neutrino physics would add to the global effort.  Charge and light readout methodologies can be applicable to both efforts.

\subsection{Gaseous TPCs for Directional Recoil Detection}

TPCs with moderate-density target gases and highly segmented readouts, typically in the form of micro-pattern gaseous detectors (MPGDs), are capable of detecting the direction of low-energy nuclear and electronic recoils in a large variety of gases~\cite{Vahsen:2021gnb}. This should enable directional CE$\nu$NS measurements for a range of nuclei, which is interesting for the core COHERENT program, and also allows novel searches for BSM MeV-scale mediators, which can modify the angular distributions in CE$\nu$NS events~\cite{Abdullah:2020iiv}. The directional detection of CE$\nu$NS would also demonstrate the proposed strategy of separating neutrino and DM-scattering events via directionality in a large observatory that probes solar neutrinos and DM simultaneously, in the so-called neutrino fog~\cite{Vahsen:2020pzb}.

\begin{figure}[ht!]
\begin{center}
\includegraphics[width=0.52\columnwidth]{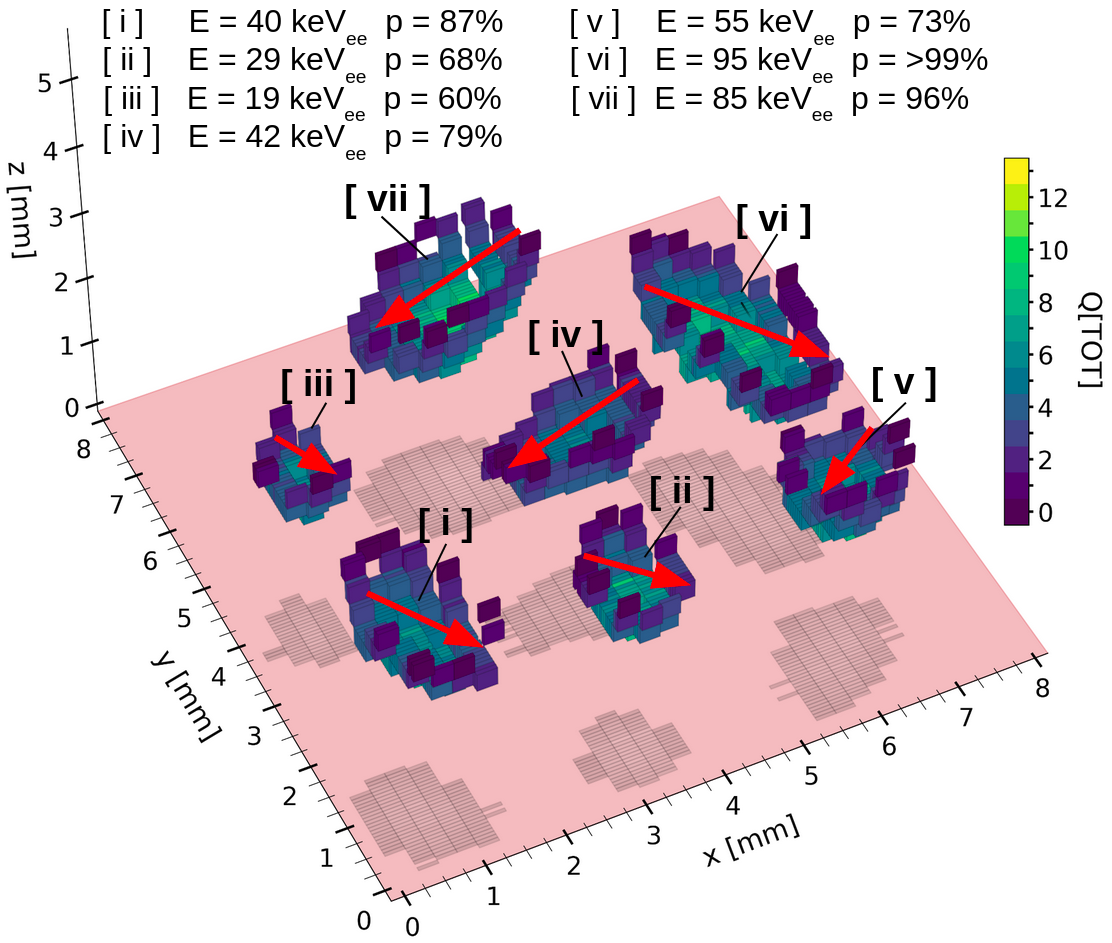}
\includegraphics[width=0.46\columnwidth]{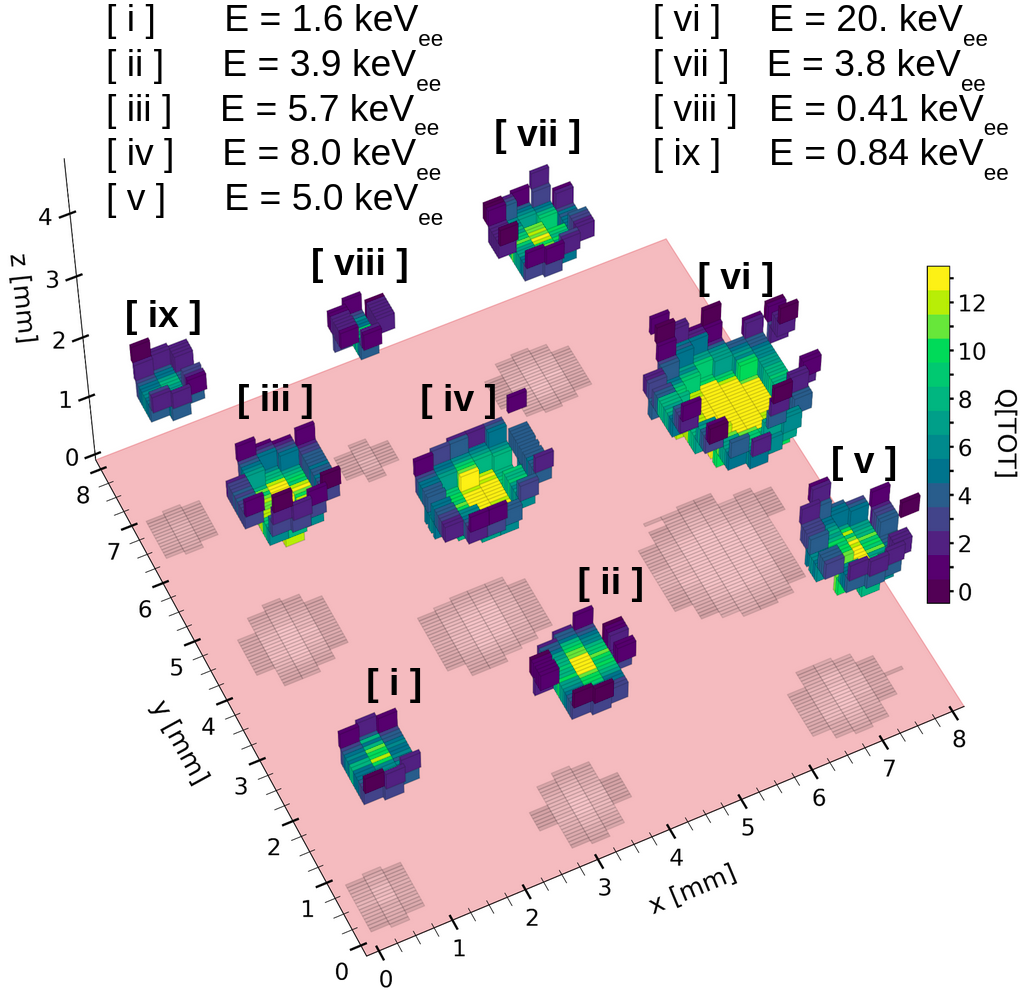}
\caption{From ~\cite{OHare:2022jnx}. Helium recoils events detected in a gas TPC with GEM amplification and pixel ASIC charge readout~\cite{Jaegle:2019jpx} in atmospheric-pressure He:CO$_2$ (70:30) gas. The 3D voxel size is $50 \times 250 \times 250~\mu$m$^3$, and raw data is shown, without any post-processing. The color scale indicates ionization density. Left: Low-gain operation (gain $900$). The red arrows show fitted recoil directions, with the head and tail (i.e., sign of the vectors) determined by a 3D convolutional neural network (3DCNN). The confidence level of correct assignment is indicated in the legend. Right: Recoils observed in same detector, but now operating in the single-electron detection regime, at a gain of $1.3\times 10^4$. In this case even sub-keV recoils are easily detected above noise. While the recoil direction is no longer visible to the eye, in simulation a 3DCNN is capable of assigning head-tail directions to energies as low as $3~{\rm keVee}$.}\label{gas_tpc_events}
\end{center}
\end{figure}

Depending on the exact detector configuration and fill gas, directional TPC detectors can detect even single electrons of ionization, and reconstruction of the event topology and recoil direction (feasible when the recoil length significantly exceeds the typical electron diffusion length during drift of ionization in the detector) is achievable for ionization energies as low as tens or even just a few keV---see Fig.~\ref{gas_tpc_events}. Significant further performance improvements may be achievable, for example with negative-ion-drift gases~\cite{Phan:2016veo} and/or lower gas densities.

Because of a strong trade-off between target mass and the minimum energy where directionality is available (both increase with gas density), a realistic directional CEvNS detector at SNS would need more than $1{\rm m}^3$ of target volume to see interesting annual event rates with a keV-scale directionality threshold at FTS, while space is very limited. This makes the STS, with increased neutrino rates and more space for detectors, optimal for directional gas TPCs. Much smaller detectors~\cite{Jaegle:2019jpx} are already available for installation. These would enable a first directional investigation of the neutron background and of the event timing capabilities, which are required to take advantage of coincidence with the SNS beam pulses. Achieving optimal event timing will require correcting for the TPC drift time by reconstructing the event coordinate in the drift direction. This coordinate reconstruction has already been demonstrated both in electron and negative ion drift gases~\cite{Lewis:2021mgp,Phan:2016veo}.

\section{Facilities at the SNS}
\label{sec:facilities}

\begin{figure}[htbp]\centering
    \includegraphics[width=\textwidth]{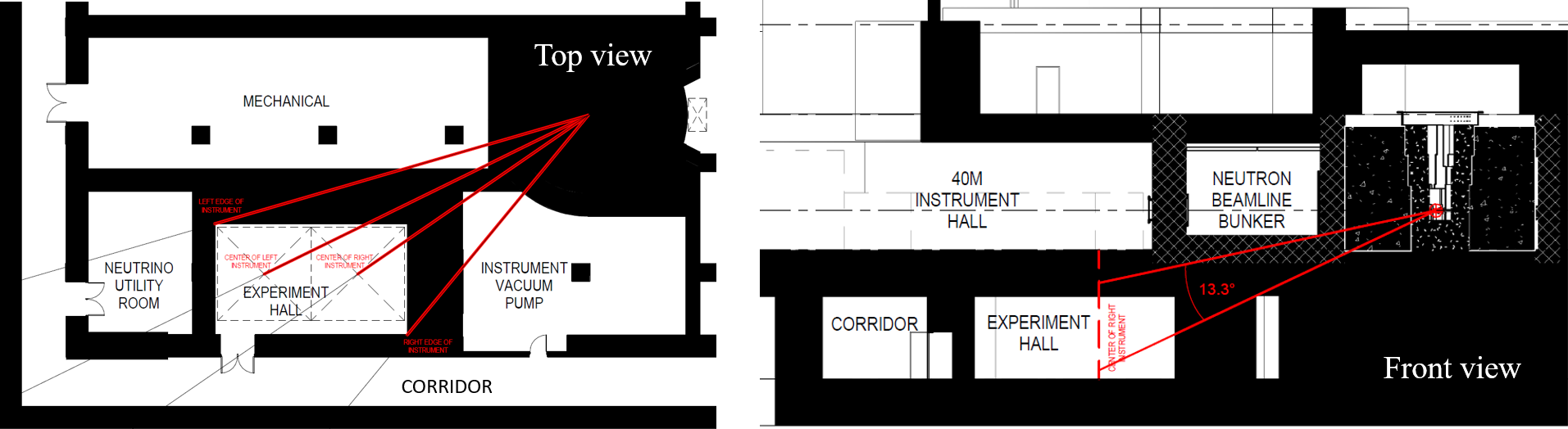}
    \caption{Top and front (look back to the beam) views of the planned neutrino experimental hall and utility room at the STS.}
    \label{f:hallviews}
\end{figure}

Specific detectors may have specific needs, but generically,
a hall of 4.5~m $\times$ 10~m $\times$~4~m height would serve for a 10-tonne scale detector.
For most physics topics, one wants to be as close as possible to the target, with as little background as possible; there is some tradeoff between proximity to the neutrino source and adequate shielding against neutrons. For some physics for which the signal is baseline-dependent, such as for sterile oscillations, specific locations may be desired.  For accelerator-produced dark matter searches, angle with respect to the beam axis is important.  In these latter cases, detector movability is also desirable.
 All of these are feasible at the STS for relatively modest investment.  Figure~\ref{f:hallviews} shows some examples.
We note that in some cases additional investment in facilities at the FTS may be desirable.

\section{Summary}
\label{sec:summary}

In this white paper we highlight several opportunities of broad relevance for particle physics and for detector development at the upgraded SNS.  Neutrino facilities at the STS will be necessary to fully exploit the high-quality neutrino source.

\newpage

\bibliographystyle{JHEP}
\bibliography{refs.bib}

\providecommand{\href}[2]{#2}\begingroup\raggedright\begin{thebibliography}{10}

\bibitem{coherentwp}
D.~Akimov et~al., \emph{{The COHERENT Experimental Program}},  in \emph{{2022
  Snowmass Summer Study}}, 4, 2022
  [\href{https://arxiv.org/abs/2204.04575}{{\ttfamily arXiv:2204.04575}}].

\bibitem{COHERENT:2017ipa}
{\scshape COHERENT} collaboration, \emph{{Observation of Coherent Elastic
  Neutrino-Nucleus Scattering}},
  \href{https://doi.org/10.1126/science.aao0990}{\emph{Science} {\bfseries 357}
  (2017) 1123} [\href{https://arxiv.org/abs/1708.01294}{{\ttfamily
  arXiv:1708.01294}}].

\bibitem{COHERENT:2020iec}
{\scshape COHERENT} collaboration, \emph{{First Measurement of Coherent Elastic
  Neutrino-Nucleus Scattering on Argon}},
  \href{https://doi.org/10.1103/PhysRevLett.126.012002}{\emph{Phys. Rev. Lett.}
  {\bfseries 126} (2021) 012002}
  [\href{https://arxiv.org/abs/2003.10630}{{\ttfamily arXiv:2003.10630}}].

\bibitem{herwig}
K.~Herwig,
  \emph{\url{https://conference.sns.gov/event/171/attachments/258/1364/Herwig_FundPhys_Workshop.pptx}},

\bibitem{COHERENT:2021pvd}
{\scshape COHERENT} collaboration, \emph{{First Probe of Sub-GeV Dark Matter
  Beyond the Cosmological Expectation with the COHERENT CsI Detector at the
  SNS}},  \href{https://arxiv.org/abs/2110.11453}{{\ttfamily
  arXiv:2110.11453}}.

\bibitem{COHERENT:2022pli}
{\scshape COHERENT} collaboration, \emph{{A COHERENT constraint on leptophobic
  dark matter using CsI data}},
  \href{https://arxiv.org/abs/2205.12414}{{\ttfamily arXiv:2205.12414}}.

\bibitem{COHERENT:2021yvp}
{\scshape COHERENT} collaboration, \emph{{Simulating the neutrino flux from the
  Spallation Neutron Source for the COHERENT experiment}},
  \href{https://doi.org/10.1103/PhysRevD.106.032003}{\emph{Phys. Rev. D}
  {\bfseries 106} (2022) 032003}
  [\href{https://arxiv.org/abs/2109.11049}{{\ttfamily arXiv:2109.11049}}].

\bibitem{Bolozdynya:2012xv}
A.~Bolozdynya et~al., \emph{{Opportunities for Neutrino Physics at the
  Spallation Neutron Source: A White Paper}},  11, 2012
  [\href{https://arxiv.org/abs/1211.5199}{{\ttfamily arXiv:1211.5199}}].

\bibitem{Freedman:1973yd}
D.Z.~Freedman, \emph{{Coherent Neutrino Nucleus Scattering as a Probe of the
  Weak Neutral Current}},
  \href{https://doi.org/10.1103/PhysRevD.9.1389}{\emph{Phys. Rev. D} {\bfseries
  9} (1974) 1389}.

\bibitem{Kopeliovich:1974mv}
V.B.~Kopeliovich and L.L.~Frankfurt, \emph{{Isotopic and chiral structure of
  neutral current}}, {\emph{JETP Lett.} {\bfseries 19} (1974) 145}.

\bibitem{Abdullah:2022zue}
M.~Abdullah et~al., \emph{{Coherent elastic neutrino-nucleus scattering:
  Terrestrial and astrophysical applications}},
  \href{https://arxiv.org/abs/2203.07361}{{\ttfamily arXiv:2203.07361}}.

\bibitem{Garvey:2005pn}
G.T.~Garvey, A.~Green, C.~Green, W.C.~Louis, G.B.~Mills, G.~McGregor et~al.,
  \emph{{Measuring active-sterile neutrino oscillations with a stopped pion
  neutrino source}},
  \href{https://doi.org/10.1103/PhysRevD.72.092001}{\emph{Phys. Rev. D}
  {\bfseries 72} (2005) 092001}
  [\href{https://arxiv.org/abs/hep-ph/0501013}{{\ttfamily
  arXiv:hep-ph/0501013}}].

\bibitem{Anderson:2012pn}
A.J.~Anderson, J.M.~Conrad, E.~Figueroa-Feliciano, C.~Ignarra, G.~Karagiorgi,
  K.~Scholberg et~al., \emph{{Measuring Active-to-Sterile Neutrino Oscillations
  with Neutral Current Coherent Neutrino-Nucleus Scattering}},
  \href{https://doi.org/10.1103/PhysRevD.86.013004}{\emph{Phys. Rev. D}
  {\bfseries 86} (2012) 013004}
  [\href{https://arxiv.org/abs/1201.3805}{{\ttfamily arXiv:1201.3805}}].

\bibitem{Barbeau:2021exu}
P.S.~Barbeau, Y.~Efremenko and K.~Scholberg, \emph{{COHERENT at the Spallation
  Neutron Source}},  \href{https://arxiv.org/abs/2111.07033}{{\ttfamily
  arXiv:2111.07033}}.

\bibitem{PhysRevD.70.023514}
P.~Fayet, \emph{Light spin-$\frac{1}{2}$ or spin-0 dark matter particles},
  \href{https://doi.org/10.1103/PhysRevD.70.023514}{\emph{Phys. Rev. D}
  {\bfseries 70} (2004) 023514}.

\bibitem{BOEHM2004219}
C.~Bœhm and P.~Fayet, \emph{Scalar dark matter candidates},
  \href{https://doi.org/https://doi.org/10.1016/j.nuclphysb.2004.01.015}{\emph{Nuclear
  Physics B} {\bfseries 683} (2004) 219}.

\bibitem{Pospelov:2007mp}
M.~Pospelov, A.~Ritz and M.B.~Voloshin, \emph{{Secluded WIMP Dark Matter}},
  \href{https://doi.org/10.1016/j.physletb.2008.02.052}{\emph{Phys. Lett. B}
  {\bfseries 662} (2008) 53} [\href{https://arxiv.org/abs/0711.4866}{{\ttfamily
  arXiv:0711.4866}}].

\bibitem{PhysRevD.84.075020}
P.~deNiverville, M.~Pospelov and A.~Ritz, \emph{Observing a light dark matter
  beam with neutrino experiments},
  \href{https://doi.org/10.1103/PhysRevD.84.075020}{\emph{Phys. Rev. D}
  {\bfseries 84} (2011) 075020}.

\bibitem{Aranda:1998fr}
A.~Aranda and C.D.~Carone, \emph{{Limits on a light leptophobic gauge boson}},
  \href{https://doi.org/10.1016/S0370-2693(98)01309-4}{\emph{Phys. Lett. B}
  {\bfseries 443} (1998) 352}
  [\href{https://arxiv.org/abs/hep-ph/9809522}{{\ttfamily
  arXiv:hep-ph/9809522}}].

\bibitem{Gondolo2012LightDM}
P.~Gondolo, P.~Ko and Y.~Omura, \emph{Light dark matter in leptophobic z'
  models}, {\emph{Physical Review D} {\bfseries 85} (2012) 035022}.

\bibitem{Batell:2014yra}
B.~Batell, P.~deNiverville, D.~McKeen, M.~Pospelov and A.~Ritz,
  \emph{{Leptophobic Dark Matter at Neutrino Factories}},
  \href{https://doi.org/10.1103/PhysRevD.90.115014}{\emph{Phys. Rev. D}
  {\bfseries 90} (2014) 115014}
  [\href{https://arxiv.org/abs/1405.7049}{{\ttfamily arXiv:1405.7049}}].

\bibitem{deNiverville:2015mwa}
P.~deNiverville, M.~Pospelov and A.~Ritz, \emph{{Light new physics in coherent
  neutrino-nucleus scattering experiments}},
  \href{https://doi.org/10.1103/PhysRevD.92.095005}{\emph{Phys. Rev. D}
  {\bfseries 92} (2015) 095005}
  [\href{https://arxiv.org/abs/1505.07805}{{\ttfamily arXiv:1505.07805}}].

\bibitem{deNiverville:2016rqh}
P.~deNiverville, C.-Y.~Chen, M.~Pospelov and A.~Ritz, \emph{{Light dark matter
  in neutrino beams: production modelling and scattering signatures at
  MiniBooNE, T2K and SHiP}},
  \href{https://doi.org/10.1103/PhysRevD.95.035006}{\emph{Phys. Rev.}
  {\bfseries D95} (2017) 035006}
  [\href{https://arxiv.org/abs/1609.01770}{{\ttfamily arXiv:1609.01770}}].

\bibitem{Boyarsky:2021moj}
A.~Boyarsky, O.~Mikulenko, M.~Ovchynnikov and L.~Shchutska, \emph{{Searches for
  new physics at SND@LHC}},  \href{https://arxiv.org/abs/2104.09688}{{\ttfamily
  arXiv:2104.09688}}.

\bibitem{Dutta:2022tav}
B.~Dutta, W.-C.~Huang, J.L.~Newstead and V.~Pandey, \emph{{Inelastic nuclear
  scattering from neutrinos and dark matter}},
  \href{https://arxiv.org/abs/2206.08590}{{\ttfamily arXiv:2206.08590}}.

\bibitem{Batell:2022xau}
B.~Batell et~al., \emph{{Dark Sector Studies with Neutrino Beams}},  in
  \emph{{2022 Snowmass Summer Study}}, 7, 2022
  [\href{https://arxiv.org/abs/2207.06898}{{\ttfamily arXiv:2207.06898}}].

\bibitem{Capozzi:2018dat}
F.~Capozzi, S.W.~Li, G.~Zhu and J.F.~Beacom, \emph{{DUNE as the Next-Generation
  Solar Neutrino Experiment}},
  \href{https://doi.org/10.1103/PhysRevLett.123.131803}{\emph{Phys. Rev. Lett.}
  {\bfseries 123} (2019) 131803}
  [\href{https://arxiv.org/abs/1808.08232}{{\ttfamily arXiv:1808.08232}}].

\bibitem{Abi:2021vq}
B.~Abi, R.~Acciarri, M.A.~Acero, G.~Adamov, D.~Adams, M.~Adinolfi et~al.,
  \emph{Supernova neutrino burst detection with the deep underground neutrino
  experiment}, \href{https://doi.org/10.1140/epjc/s10052-021-09166-w}{\emph{The
  European Physical Journal C} {\bfseries 81} (2021) 423}.

\bibitem{LSND:2001aii}
{\scshape LSND} collaboration, \emph{{Evidence for neutrino oscillations from
  the observation of $\bar{\nu}_e$ appearance in a $\bar{\nu}_\mu$ beam}},
  \href{https://doi.org/10.1103/PhysRevD.64.112007}{\emph{Phys. Rev. D}
  {\bfseries 64} (2001) 112007}
  [\href{https://arxiv.org/abs/hep-ex/0104049}{{\ttfamily
  arXiv:hep-ex/0104049}}].

\bibitem{MiniBooNE:2020pnu}
{\scshape MiniBooNE} collaboration, \emph{{Updated MiniBooNE neutrino
  oscillation results with increased data and new background studies}},
  \href{https://doi.org/10.1103/PhysRevD.103.052002}{\emph{Phys. Rev. D}
  {\bfseries 103} (2021) 052002}
  [\href{https://arxiv.org/abs/2006.16883}{{\ttfamily arXiv:2006.16883}}].

\bibitem{Mention:2011rk}
G.~Mention, M.~Fechner, T.~Lasserre, T.A.~Mueller, D.~Lhuillier, M.~Cribier
  et~al., \emph{{The Reactor Antineutrino Anomaly}},
  \href{https://doi.org/10.1103/PhysRevD.83.073006}{\emph{Phys. Rev. D}
  {\bfseries 83} (2011) 073006}
  [\href{https://arxiv.org/abs/1101.2755}{{\ttfamily arXiv:1101.2755}}].

\bibitem{SAGE:1998fvr}
{\scshape SAGE} collaboration, \emph{{Measurement of the response of the
  Russian-American gallium experiment to neutrinos from a Cr-51 source}},
  \href{https://doi.org/10.1103/PhysRevC.59.2246}{\emph{Phys. Rev. C}
  {\bfseries 59} (1999) 2246}
  [\href{https://arxiv.org/abs/hep-ph/9803418}{{\ttfamily
  arXiv:hep-ph/9803418}}].

\bibitem{Abdurashitov:2005tb}
J.N.~Abdurashitov et~al., \emph{{Measurement of the response of a Ga solar
  neutrino experiment to neutrinos from an Ar-37 source}},
  \href{https://doi.org/10.1103/PhysRevC.73.045805}{\emph{Phys. Rev. C}
  {\bfseries 73} (2006) 045805}
  [\href{https://arxiv.org/abs/nucl-ex/0512041}{{\ttfamily
  arXiv:nucl-ex/0512041}}].

\bibitem{Kaether:2010ag}
F.~Kaether, W.~Hampel, G.~Heusser, J.~Kiko and T.~Kirsten, \emph{{Reanalysis of
  the GALLEX solar neutrino flux and source experiments}},
  \href{https://doi.org/10.1016/j.physletb.2010.01.030}{\emph{Phys. Lett. B}
  {\bfseries 685} (2010) 47} [\href{https://arxiv.org/abs/1001.2731}{{\ttfamily
  arXiv:1001.2731}}].

\bibitem{Gariazzo:2017fdh}
S.~Gariazzo, C.~Giunti, M.~Laveder and Y.F.~Li, \emph{{Updated Global 3+1
  Analysis of Short-BaseLine Neutrino Oscillations}},
  \href{https://doi.org/10.1007/JHEP06(2017)135}{\emph{JHEP} {\bfseries 06}
  (2017) 135} [\href{https://arxiv.org/abs/1703.00860}{{\ttfamily
  arXiv:1703.00860}}].

\bibitem{COHERENT:2021xhx}
{\scshape COHERENT} collaboration, \emph{{A D$_2$O detector for flux
  normalization of a pion decay-at-rest neutrino source}},
  \href{https://doi.org/10.1088/1748-0221/16/08/P08048}{\emph{JINST} {\bfseries
  16} (2021) P08048} [\href{https://arxiv.org/abs/2104.09605}{{\ttfamily
  arXiv:2104.09605}}].

\bibitem{COHERENT:2018gft}
{\scshape COHERENT} collaboration, \emph{{COHERENT 2018 at the Spallation
  Neutron Source}},  \href{https://arxiv.org/abs/1803.09183}{{\ttfamily
  arXiv:1803.09183}}.

\bibitem{Alexander:2019uvv}
T.~Alexander et~al., \emph{{The Low-Radioactivity Underground Argon Workshop: A
  workshop synopsis}},  in \emph{{Low-Radioactivity Underground Argon Richland,
  Washington, USA, March 19-20, 2018}}, 2019
  [\href{https://arxiv.org/abs/1901.10108}{{\ttfamily arXiv:1901.10108}}].

\bibitem{Akimov:2019eae}
D.~Akimov, V.~Belov, A.~Konovalov, A.~Kumpan, O.~Razuvaeva, D.~Rudik et~al.,
  \emph{{Fast component re-emission in Xe-doped liquid argon}},
  \href{https://doi.org/10.1088/1748-0221/14/09/P09022}{\emph{JINST} {\bfseries
  14} (2019) P09022} [\href{https://arxiv.org/abs/1906.00836}{{\ttfamily
  arXiv:1906.00836}}].

\bibitem{Caratelli:2022llt}
D.~Caratelli et~al., \emph{{Low-Energy Physics in Neutrino LArTPCs}},
  \href{https://arxiv.org/abs/2203.00740}{{\ttfamily arXiv:2203.00740}}.

\bibitem{Anderson:2012vc}
C.~Anderson, M.~Antonello, B.~Baller, T.~Bolton, C.~Bromberg, F.~Cavanna
  et~al., \emph{{The ArgoNeuT Detector in the NuMI Low-Energy beam line at
  Fermilab}}, \href{https://doi.org/10.1088/1748-0221/7/10/P10019}{\emph{JINST}
  {\bfseries 7} (2012) P10019}
  [\href{https://arxiv.org/abs/1205.6747}{{\ttfamily arXiv:1205.6747}}].

\bibitem{LArIAT:2019kzd}
{\scshape LArIAT Collaboration} collaboration, \emph{{The Liquid Argon In A
  Testbeam (LArIAT) Experiment}},
  \href{https://doi.org/10.1088/1748-0221/15/04/P04026}{\emph{JINST} {\bfseries
  15} (2020) P04026} [\href{https://arxiv.org/abs/1911.10379}{{\ttfamily
  arXiv:1911.10379}}].

\bibitem{Bian:2015qka}
J.~Bian, \emph{{The CAPTAIN Experiment}},  in \emph{{Meeting of the APS
  Division of Particles and Fields}}, 9, 2015
  [\href{https://arxiv.org/abs/1509.07739}{{\ttfamily arXiv:1509.07739}}].

\bibitem{antonello2015operation}
M.~Antonello, P.~Aprili, B.~Baibussinov, F.~Boffelli, A.~Bubak, E.~Calligarich
  et~al., \emph{{Operation and performance of the ICARUS-T600 cryogenic plant
  at Gran Sasso underground Laboratory}},
  \href{https://doi.org/10.1088/1748-0221/10/12/P12004}{\emph{JINST} {\bfseries
  10} (2015) P12004} [\href{https://arxiv.org/abs/1504.01556}{{\ttfamily
  arXiv:1504.01556}}].

\bibitem{acciarri2017design}
{\scshape MicroBooNE Collaboration} collaboration, \emph{{Design and
  Construction of the MicroBooNE Detector}},
  \href{https://doi.org/10.1088/1748-0221/12/02/P02017}{\emph{JINST} {\bfseries
  12} (2017) P02017} [\href{https://arxiv.org/abs/1612.05824}{{\ttfamily
  arXiv:1612.05824}}].

\bibitem{Qian:2018qbv}
X.~Qian, C.~Zhang, B.~Viren and M.~Diwan, \emph{{Three-dimensional Imaging for
  Large LArTPCs}},
  \href{https://doi.org/10.1088/1748-0221/13/05/P05032}{\emph{JINST} {\bfseries
  13} (2018) P05032} [\href{https://arxiv.org/abs/1803.04850}{{\ttfamily
  arXiv:1803.04850}}].

\bibitem{MicroBooNE:2020vry}
{\scshape MicroBooNE Collaboration} collaboration, \emph{{Neutrino event
  selection in the MicroBooNE liquid argon time projection chamber using
  Wire-Cell 3D imaging, clustering, and charge-light matching}},
  \href{https://doi.org/10.1088/1748-0221/16/06/P06043}{\emph{JINST} {\bfseries
  16} (2021) P06043} [\href{https://arxiv.org/abs/2011.01375}{{\ttfamily
  arXiv:2011.01375}}].

\bibitem{MicroBooNE:2020jgj}
{\scshape MicroBooNE Collaboration} collaboration, \emph{{High-performance
  Generic Neutrino Detection in a LArTPC near the Earth's Surface with the
  MicroBooNE Detector}},  \href{https://arxiv.org/abs/2012.07928}{{\ttfamily
  arXiv:2012.07928}}.

\bibitem{MicroBooNE:2021zul}
{\scshape MicroBooNE Collaboration} collaboration, \emph{{Cosmic Ray Background
  Rejection with Wire-Cell LArTPC Event Reconstruction in the MicroBooNE
  Detector}},
  \href{https://doi.org/10.1103/PhysRevApplied.15.064071}{\emph{Phys. Rev.
  Applied} {\bfseries 15} (2021) 064071}
  [\href{https://arxiv.org/abs/2101.05076}{{\ttfamily arXiv:2101.05076}}].

\bibitem{MicroBooNE:2021ojx}
{\scshape MicroBooNE Collaboration} collaboration, \emph{{Wire-Cell 3D Pattern
  Recognition Techniques for Neutrino Event Reconstruction in Large LArTPCs:
  Algorithm Description and Quantitative Evaluation with MicroBooNE
  Simulation}},
  \href{https://doi.org/10.1088/1748-0221/17/01/P01037}{\emph{JINST} {\bfseries
  17} (2022) P01037} [\href{https://arxiv.org/abs/2110.13961}{{\ttfamily
  arXiv:2110.13961}}].

\bibitem{MicroBooNE:2017qiu}
{\scshape MicroBooNE Collaboration} collaboration, \emph{{Noise
  Characterization and Filtering in the MicroBooNE Liquid Argon TPC}},
  \href{https://doi.org/10.1088/1748-0221/12/08/P08003}{\emph{JINST} {\bfseries
  12} (2017) P08003} [\href{https://arxiv.org/abs/1705.07341}{{\ttfamily
  arXiv:1705.07341}}].

\bibitem{LArIAT:2019gdz}
{\scshape LArIAT Collaboration} collaboration, \emph{{Calorimetry for
  low-energy electrons using charge and light in liquid argon}},
  \href{https://doi.org/10.1103/PhysRevD.101.012010}{\emph{Phys. Rev. D}
  {\bfseries 101} (2020) 012010}
  [\href{https://arxiv.org/abs/1909.07920}{{\ttfamily arXiv:1909.07920}}].

\bibitem{Dwyer:2018phu}
D.A.~Dwyer, M.~Garcia-Sciveres, D.~Gnani, C.~Grace, S.~Kohn, M.~Kramer et~al.,
  \emph{{LArPix: Demonstration of low-power 3D pixelated charge readout for
  liquid argon time projection chambers}},
  \href{https://doi.org/10.1088/1748-0221/13/10/P10007}{\emph{JINST} {\bfseries
  13} (2018) P10007} [\href{https://arxiv.org/abs/1808.02969}{{\ttfamily
  arXiv:1808.02969}}].

\bibitem{Nygren:2018rbl}
D.~Nygren and Y.~Mei, \emph{{Q-Pix: Pixel-scale Signal Capture for Kiloton
  Liquid Argon TPC Detectors: Time-to-Charge Waveform Capture, Local Clocks,
  Dynamic Networks}},  \href{https://arxiv.org/abs/1809.10213}{{\ttfamily
  arXiv:1809.10213}}.

\bibitem{Adams:2019uqx}
C.~Adams, M.~Del~Tutto, J.~Asaadi, M.~Bernstein, E.~Church, R.~Guenette et~al.,
  \emph{{Enhancing neutrino event reconstruction with pixel-based 3D readout
  for liquid argon time projection chambers}},
  \href{https://doi.org/10.1088/1748-0221/15/04/P04009}{\emph{JINST} {\bfseries
  15} (2020) P04009} [\href{https://arxiv.org/abs/1912.10133}{{\ttfamily
  arXiv:1912.10133}}].

\bibitem{Q-Pix:2022zjm}
{\scshape Q-Pix} collaboration, \emph{{Enhanced low-energy supernova burst
  detection in large liquid argon time projection chambers enabled by Q-Pix}},
  \href{https://doi.org/10.1103/PhysRevD.106.032011}{\emph{Phys. Rev. D}
  {\bfseries 106} (2022) 032011}
  [\href{https://arxiv.org/abs/2203.12109}{{\ttfamily arXiv:2203.12109}}].

\bibitem{roberto_mandujano_2022_6805002}
R.~Mandujano, \emph{Dune nd-lar: Design and status},  July, 2022.
\newblock 10.5281/zenodo.6805002.

\bibitem{PhysRevD.99.012002}
{\scshape ArgoNeuT Collaboration} collaboration, \emph{Demonstration of
  mev-scale physics in liquid argon time projection chambers using argoneut},
  \href{https://doi.org/10.1103/PhysRevD.99.012002}{\emph{Phys. Rev. D}
  {\bfseries 99} (2019) 012002}.

\bibitem{1748-0221-12-09-P09014}
R.~Acciarri, C.~Adams, R.~An, J.~Anthony, J.~Asaadi, M.~Auger et~al.,
  \emph{Michel electron reconstruction using cosmic-ray data from the
  microboone lartpc}, {\emph{Journal of Instrumentation} {\bfseries 12} (2017)
  P09014}.

\bibitem{instruments5040031}
A.A.~Abud, B.~Abi, R.~Acciarri, M.A.~Acero, G.~Adamov, D.~Adams et~al.,
  \emph{Deep underground neutrino experiment (dune) near detector conceptual
  design report},
  \href{https://doi.org/10.3390/instruments5040031}{\emph{Instruments}
  {\bfseries 5} (2021) }.

\bibitem{https://doi.org/10.48550/arxiv.2203.06894}
T.~Aramaki, M.~Boezio, J.~Buckley, E.~Bulbul, P.~von Doetinchem, F.~Donato
  et~al., \emph{Snowmass2021 cosmic frontier: The landscape of cosmic-ray and
  high-energy photon probes of particle dark matter},  2022.
\newblock 10.48550/ARXIV.2203.06894.

\bibitem{PhysRevD.99.036009}
A.~Friedland and S.W.~Li, \emph{Understanding the energy resolution of liquid
  argon neutrino detectors},
  \href{https://doi.org/10.1103/PhysRevD.99.036009}{\emph{Phys. Rev. D}
  {\bfseries 99} (2019) 036009}.

\bibitem{Rooks:2022vrl}
M.~Rooks, S.~Abbaszadeh, J.~Asaadi, M.~Febbraro, R.W.~Gladen, E.~Gramellini
  et~al., \emph{{Development of a novel, windowless, amorphous selenium based
  photodetector for use in liquid noble detectors}},
  \href{https://arxiv.org/abs/2207.11127}{{\ttfamily arXiv:2207.11127}}.

\bibitem{instruments5010004}
M.~Ku{\'z}niak and A.M.~Szelc, \emph{Wavelength shifters for applications in
  liquid argon detectors},
  \href{https://doi.org/10.3390/instruments5010004}{\emph{Instruments}
  {\bfseries 5} (2021) }.

\bibitem{instruments2010003}
M.~Auger, Y.~Chen, A.~Ereditato, D.~Goeldi, I.~Kreslo, D.~Lorca et~al.,
  \emph{Arclight---a compact dielectric large-area photon detector},
  \href{https://doi.org/10.3390/instruments2010003}{\emph{Instruments}
  {\bfseries 2} (2018) }.

\bibitem{Anfimov_2020}
N.~Anfimov, R.~Berner, I.~Butorov, A.~Chetverikov, D.~Fedoseev, B.~Gromov
  et~al., \emph{Development of the light collection module for the liquid argon
  time projection chamber ({LArTPC})},
  \href{https://doi.org/10.1088/1748-0221/15/07/c07022}{\emph{Journal of
  Instrumentation} {\bfseries 15} (2020) C07022}.

\bibitem{MicroBooNE:2019lta}
{\scshape MicroBooNE} collaboration, \emph{{Design and construction of the
  MicroBooNE Cosmic Ray Tagger system}},
  \href{https://doi.org/10.1088/1748-0221/14/04/P04004}{\emph{JINST} {\bfseries
  14} (2019) P04004} [\href{https://arxiv.org/abs/1901.02862}{{\ttfamily
  arXiv:1901.02862}}].

\bibitem{Machado:2019oxb}
P.A.~Machado, O.~Palamara and D.W.~Schmitz, \emph{{The Short-Baseline Neutrino
  Program at Fermilab}},
  \href{https://doi.org/10.1146/annurev-nucl-101917-020949}{\emph{Ann. Rev.
  Nucl. Part. Sci.} {\bfseries 69} (2019) 363}
  [\href{https://arxiv.org/abs/1903.04608}{{\ttfamily arXiv:1903.04608}}].

\bibitem{Akimov:2021dab}
{\scshape COHERENT} collaboration, \emph{{Measurement of the Coherent Elastic
  Neutrino-Nucleus Scattering Cross Section on CsI by COHERENT}},
  \href{https://arxiv.org/abs/2110.07730}{{\ttfamily arXiv:2110.07730}}.

\bibitem{Mosconi:2007tz}
B.~Mosconi, P.~Ricci, E.~Truhlik and P.~Vogel, \emph{{Model dependence of the
  neutrino-deuteron disintegration cross sections at low energies}},
  \href{https://doi.org/10.1103/PhysRevC.75.044610}{\emph{Phys. Rev. C}
  {\bfseries 75} (2007) 044610}
  [\href{https://arxiv.org/abs/nucl-th/0702073}{{\ttfamily
  arXiv:nucl-th/0702073}}].

\bibitem{Ando:2019yum}
S.-I.~Ando, Y.-H.~Song and C.H.~Hyun, \emph{{Neutrino-Deuteron Reactions at
  Solar Neutrino Energies in Pionless Effective Field Theory with Dibaryon
  Fields}}, \href{https://doi.org/10.1103/PhysRevC.101.054001}{\emph{Phys. Rev.
  C} {\bfseries 101} (2020) 054001}
  [\href{https://arxiv.org/abs/1911.11307}{{\ttfamily arXiv:1911.11307}}].

\bibitem{Super-Kamiokande:2017yvm}
{\scshape Super-Kamiokande} collaboration, \emph{{Atmospheric neutrino
  oscillation analysis with external constraints in Super-Kamiokande I-IV}},
  \href{https://doi.org/10.1103/PhysRevD.97.072001}{\emph{Phys. Rev. D}
  {\bfseries 97} (2018) 072001}
  [\href{https://arxiv.org/abs/1710.09126}{{\ttfamily arXiv:1710.09126}}].

\bibitem{Hyper-Kamiokande:2018ofw}
{\scshape Hyper-Kamiokande} collaboration, \emph{{Hyper-Kamiokande Design
  Report}},  \href{https://arxiv.org/abs/1805.04163}{{\ttfamily
  arXiv:1805.04163}}.

\bibitem{Vahsen:2021gnb}
S.E.~Vahsen, C.A.J.~O'Hare and D.~Loomba, \emph{{Directional Recoil
  Detection}},
  \href{https://doi.org/10.1146/annurev-nucl-020821-035016}{\emph{Ann. Rev.
  Nucl. Part. Sci.} {\bfseries 71} (2021) 189}
  [\href{https://arxiv.org/abs/2102.04596}{{\ttfamily arXiv:2102.04596}}].

\bibitem{Abdullah:2020iiv}
M.~Abdullah, D.~Aristizabal~Sierra, B.~Dutta and L.E.~Strigari, \emph{{Coherent
  Elastic Neutrino-Nucleus Scattering with directional detectors}},
  \href{https://doi.org/10.1103/PhysRevD.102.015009}{\emph{Phys. Rev. D}
  {\bfseries 102} (2020) 015009}
  [\href{https://arxiv.org/abs/2003.11510}{{\ttfamily arXiv:2003.11510}}].

\bibitem{Vahsen:2020pzb}
S.E.~Vahsen et~al., \emph{{CYGNUS: Feasibility of a nuclear recoil observatory
  with directional sensitivity to dark matter and neutrinos}},
  \href{https://arxiv.org/abs/2008.12587}{{\ttfamily arXiv:2008.12587}}.

\bibitem{Phan:2016veo}
N.S.~Phan, R.~Lafler, R.J.~Lauer, E.R.~Lee, D.~Loomba, J.A.J.~Matthews et~al.,
  \emph{{The novel properties of SF$_6$ for directional dark matter
  experiments}},
  \href{https://doi.org/10.1088/1748-0221/12/02/P02012}{\emph{JINST} {\bfseries
  12} (2017) P02012} [\href{https://arxiv.org/abs/1609.05249}{{\ttfamily
  arXiv:1609.05249}}].

\bibitem{OHare:2022jnx}
C.A.J.~O'Hare et~al., \emph{{Recoil imaging for dark matter, neutrinos, and
  physics beyond the Standard Model}},  in \emph{{2022 Snowmass Summer Study}},
  3, 2022 [\href{https://arxiv.org/abs/2203.05914}{{\ttfamily
  arXiv:2203.05914}}].

\bibitem{Jaegle:2019jpx}
I.~Jaegle et~al., \emph{{Compact, directional neutron detectors capable of
  high-resolution nuclear recoil imaging}},
  \href{https://doi.org/10.1016/j.nima.2019.06.037}{\emph{Nucl. Instrum. Meth.
  A} {\bfseries 945} (2019) 162296}
  [\href{https://arxiv.org/abs/1901.06657}{{\ttfamily arXiv:1901.06657}}].

\bibitem{Lewis:2021mgp}
P.M.~Lewis, M.T.~Hedges, I.~Jaegle, J.~Schueler, T.N.~Thorpe and S.E.~Vahsen,
  \emph{{Primary track recovery in high-definition gas time projection
  chambers}}, \href{https://doi.org/10.1140/epjc/s10052-022-10283-3}{\emph{Eur.
  Phys. J. C} {\bfseries 82} (2022) 324}
  [\href{https://arxiv.org/abs/2106.15829}{{\ttfamily arXiv:2106.15829}}].

\end{thebibliography}\endgroup

\clearpage

\end{document}